\documentclass[12pt]{article}
\usepackage{amsmath}
\usepackage{graphicx}
\usepackage{enumerate}
\usepackage{natbib}
\usepackage{url} 

\newcommand{\blind}{1}

\usepackage{amsmath}
\usepackage{graphicx}
\usepackage{enumerate}
\usepackage{url} 

\usepackage{multicol}
\usepackage{lipsum}

\usepackage{psfrag,epsf,color,subfigure}
\usepackage{amssymb}
\usepackage{multirow}
\usepackage{array}
\usepackage{bm}
\usepackage{booktabs}
\usepackage{float}
\usepackage{caption}
\usepackage[font=small,labelfont=bf]{caption}
\setcitestyle{citesep={;}}
\usepackage{xr}

\usepackage{soul}

\usepackage{algorithm}
\usepackage{algorithmicx}
\usepackage{algpseudocode}

\newcommand{\Mean}{{\mathbb{E}}}
\newcommand{\Var}{{\mbox{Var}}}

\newtheorem{coro}{Corollary}

\newtheorem{thm}{Theorem}  
\newtheorem{lemma}{Lemma}[section]

\newtheorem{assumption}{Assumption}[section]

\begin{document}

\def\spacingset#1{\renewcommand{\baselinestretch}%
{#1}\small\normalsize} \spacingset{1}


\if1\blind
{
  \title{\bf Efficiently Learning Synthetic Control  Models for High-dimensional Disaggregated Data}
  \author{Ye Shen\thanks{Department of Statistics, North Carolina State University}, Rui Song\thanks{Amazon.com Services, Inc.},
    and Alberto Abadie\thanks{Massachusetts Institute of Technology}}
  \maketitle
} \fi

\if0\blind
{
  \bigskip
  \bigskip
  \bigskip
  \begin{center}
    {\LARGE\bf Efficiently Learning Synthetic Control  Models for High-dimensional Disaggregated Data}
\end{center}
  \medskip
} \fi

\begin{abstract}
The Synthetic Control method (SC) has become a valuable tool for estimating causal effects. Originally designed for single-treated unit scenarios, it has recently found applications in high-dimensional disaggregated settings with multiple treated units. However, challenges in practical implementation and computational efficiency arise in such scenarios. To tackle these challenges, we propose a novel approach that integrates the Multivariate Square-root Lasso method into the synthetic control framework. We rigorously establish the estimation error bounds for fitting the Synthetic Control weights using Multivariate Square-root Lasso, accommodating high-dimensionality and time series dependencies. Additionally, we quantify the estimation error for the Average Treatment Effect on the Treated (ATT). Through simulation studies, we demonstrate that our method offers superior computational efficiency without compromising estimation accuracy. We apply our method to assess the causal impact of COVID-19 Stay-at-Home Orders on the monthly unemployment rate in the United States at the county level. 
\end{abstract}

\noindent%


\spacingset{1.5} 

\section{Introduction}\label{sec:intro}

During the past decade, the Synthetic Control method \citep{abadie2003economic,abadie2010synthetic} has witnessed its increasingly wide application for the estimation of treatment effects in areas such as public health \citep{cole2020impact,  bayat2020synthetic}, crime policy \citep{robbins2017framework}, and the labor market \citep{sabia2012effects,dube2015pooling}. The classic Synthetic Control method is proposed to estimate the counterfactual outcome of a single treated unit. 
The main idea is that a weighted average of control units often provides a good approximation of the counterfactual outcome of the treated unit without treatment. To avoid extrapolation, the weights are restricted to be nonnegative and to sum to one.  Relaxation of the restrictions on the weights has been pioneered by \cite{doudchenko2016balancing}, \cite{ferman2019synthetic}, \cite{hollingsworth2020tactics}, \cite{bottmer2021design} and \cite{ben2021augmented} using regression-based methods.

While the classic Synthetic Control method is proposed for settings with a single treated unit, the method has recently found applications in settings with multiple treated units.
For example, \cite{abadie2021penalized} analyzed the impact of participation in the National Supported Work Demonstration  program on the yearly earnings in 1978 of individuals at the margins of the labor market, where there were 185 treated units and 260 control units. 
\cite{gibson2020understanding} studied the economic impact of COVID-19 stay-at-home orders on the unemployment rate with 43 treated states and 7 control states.
 Previously, \cite{kreif2016examination} evaluated the effects of a hospital P4P scheme on risk-adjusted hospital mortality with 24 treated hospitals and 132 control hospitals. \cite{robbins2017framework} investigated the effect of the Drug Market Intervention  in the Hurt Park neighborhood of Roanoke, Virginia, in late 2011 with 66 treated blocks and 3535 control blocks. And \cite{acemoglu2016value} discovered that the announcement of Timothy Geithner as the nominee for Treasury Secretary in November 2008 led to an accumulated abnormal return for 63 financial firms with which he had a previous connection, out of a total of 603 firms.

In such high-dimensional disaggregated settings, practical challenges may arise.
Firstly, due to the large number of control units, the weights used to construct the synthetic control estimator might not be unique.  
Secondly, as our simulation studies will show, fitting separate penalized Synthetic Control models iteratively for each treated unit can be time-consuming.
Existing literature suggests two possible solutions. One approach is to aggregate the treated units into a single treated unit \citep{kreif2016examination, robbins2017framework, hazlett2018trajectory}.  However, this approach has limitations, as it may generate interpolation biases and  results in the loss of individual counterfactual information, which is crucial for assessing individual treatment effects or identifying heterogeneous treatment effects in other studies \citep{agarwal2020synthetic, shen2022heterogeneous}. 
 The alternative solution is to iteratively employ penalized regression method, such as Lasso regression \citep{hollingsworth2020tactics}, restricted OLS \citep{chernozhukov2021exact} or more advanced penalty terms  \citep{abadie2021penalized}. However, as we will show in the simulation studies, \cite{hollingsworth2020tactics} does not have any theoretical guarantees and suffers from large MSE for estimating counterfactual outcomes after the treatment assignment, while \cite{abadie2021penalized} and \cite{hollingsworth2020tactics} suffer from high computational cost due to high computational complexity. So far, there has been little discussion about the efficient computation of Synthetic Control methods for multiple treated units  as outlined in \cite{abadie2021using}.

In this paper, we aim to fill the gap and provide a solution to efficiently  estimate Synthetic Control weights of multiple treated units for individual counterfactual outcome estimation. Our contributions can be summarized in four key aspects.
 
First, conceptually, we introduce a new perspective and view the problem of fitting Synthetic Control Models for multiple treated units as a Multivariate Linear Regression problem , which is, to our knowledge, the first time in the literature. This perspective opens the door to a vast body of existing literature on regression techniques.  To estimate the Synthetic Control weights efficiently, we propose to employ the Multivariate Square-root Lasso, a method known for its pivotal property and computational efficiency \citep{van2016chi,molstad2021new}.
 
Second, theoretically, we investigate the validity of fitting Synthetic Control Models using Multivariate Square-root Lasso 
 by deriving an estimation error bound for the synthetic control weights.  Our error bound is a non-trivial extension of prior results on Multivariate Square-root Lasso \citep{van2016chi, molstad2021new}  as we face unique challenges of  high-dimensionality and  time series dependency structures of the potential outcomes within the synthetic control framework. Additionally, leveraging the weight estimation error bound, we further establish an error bound for the estimation of Average Treatment Effects on the Treated (ATT).

Third, numerically, we demonstrate the empirically validity of our proposed method through extensive simulations. Our experiments illustrate a significant reduction in computation time without sacrificing estimation accuracy.  

Last but not least,  we apply our method to assess the causal impact of COVID-19 Stay-at-Home Orders on the monthly unemployment rate in the contiguous United States at the county level. 
,which evidences underscores the practicality and efficiency of our approach.

\section{Related Work} \label{sec:RelatedWork}

\textbf{Synthetic Control Methods for Multiple Treated Units.} 
In the new era of big data, there has been an increasing interest in applying Synthetic Control Methods in high-dimensional settings with multiple treated units. There are two main groups of literature. 
The first group deals with aggregated data by combining all the treated units into a single treated unit. For instance, \cite{dube2015pooling} transformed Synthetic Control estimates to elasticities, then averaged the elasticities. 
\cite{robbins2017framework} and \cite{hazlett2018trajectory} worked on the unweighted average of outcomes for all treated units.
\cite{abadie2021synthetic} propose experimental designs based on synthetic units that match aggregate feature values in the population of interest.
The second group addresses disaggregated data and emphasizes two practical challenges: the non-unique solutions for weights and overfitting concerns caused by a high-dimensional donor pool when the number of control units exceeds the number of time points. \cite{hollingsworth2020tactics} proposed a Synthetic Control Using Lasso (SCUL) that allows extrapolation and automatic donor selection. \cite{abadie2021penalized} introduced an augmented Synthetic Control estimator with a penalty term. The penalty term is weighted by the Euclidean norm of the difference between the features of the treated unit and each unit in the donor pool, which encourages the use of control units with characteristics similar to the treated unit.

\textbf{High-dimensional Multivariate Linear Regression.}
When the number of unknown parameters is greater than the number of observations, the least squares estimator is not unique. A natural alternative is a penalized least squares estimator \citep{turlach2005simultaneous, yuan2007dimension, obozinski2011support, negahban2011estimation}, which implicitly assumes that the error terms follow an identical normal distribution. Later on, to further utilize the information of the error covariance matrix, \cite{rothman2010sparse} proposed Multivariate Regression with Covariance Estimation (MRCE) to estimate the error covariance matrix and the unknown parameters jointly. MRCE maximizes a penalized normal log-likelihood by updating the error covariance matrix and the unknown parameters iteratively. Variations of MRCE was further studied by \cite{niu2019simultaneous}, \cite{chang2022robust} and \cite{molstad2021explicit}. However, sometimes we do not need the estimated error covariance matrix, and the above-mentioned methods are computationally expensive. More recently, \cite{molstad2021new} proposed the Multivariate Square-root Lasso that implicitly estimates the error covariance matrix and is computationally efficient.

 \vspace{-0.7cm}
\section{Problem Setup}\label{sec:setup}

Consider panel data with $N=m+n$ units, where the first $m$ units are treated units and the following $n$ units are control units. We assume that there are $T_0$ time points before the treatment assignment and that all treated units are treated at the same time point $T_0+1$. Without loss of generality, we assume that the time points after the treatment assignment $T_1$ equals to one. Denote $D_{i}$ as the treatment assignment indicator, where $D_{i} = 1$  if the unit $i$ is treated at time $T_0+1$, and $D_{i} = 0$ otherwise. Denote $Y_{i, t}$ as the outcome that unit $i$ receives at time point $t$.  
In  this paper, we adopt the potential outcome framework for causal inference \citep{splawa1990application,rubin1974estimating}. Specifically, let $Y_{i, t}(1)$ and $Y_{i, t}(0)$ be the potential outcome for unit $i$ in time period $t$ that would be observed if this unit receives treatment or control, respectively.

In this paper, we are interested in estimating the Average Treatment Effect on the Treated (ATT).  Denoting the vector $\mathbf{Y}_{post}$ as $\left(Y_{m+1, T_0+1},Y_{m+2, T_0+1}, \cdots, Y_{m+n, T_0+1}\right)$, we represent ATT as $\delta$ defined as follows:
\begin{equation}
\delta = \frac{1}{m}\sum_{i=1}^{m} \left\{ Y_{i,T_0+1}(1) - Y_{i,T_0+1}((0)\right\}.
\end{equation}
In essence, ATT measures the difference between the outcomes of treated units under treatment and what their outcomes would have been without treatment. This provides valuable insights into the impact of the treatment on the treated group. 
To estimate ATT, the key challenge arises from the fact that the counterfactual outcome can never be observed. In order to  establish the identification of the counterfactual outcomes, we make the following assumptions.
\begin{assumption}(No Anticipation)\label{ass:noAnticipation}
For any unit $i$ and time $t \le T_0$, we have  $Y_{i,t} = Y_{i,t}(0)$.
\end{assumption}
\vspace{-0.7cm}
\begin{assumption}(Consistency)\label{ass:Consistency} For any unit $i \in\{ 1,2,\cdots, N \}$, we have \quad
$Y_{i,T_0+1}=Y_{i,T_0+1}(1)  D_{i} + Y_{i,T_0+1}(0) (1-D_{i})$. 
\end{assumption}
Assumption \ref{ass:noAnticipation} states that the treatment has no effect on the outcome before the implementation period $T_0+1$. Assumption \ref{ass:Consistency} requires that the observed outcome of a particular unit depends only on its received treatment without the dependence on  other units' treatment assignments. Assumption \ref{ass:noAnticipation} and Assumption \ref{ass:Consistency} are both standard assumptions in causal inference literature \citep[see e.g.,][]{athey2016recursive,abadie2021using}, under which the potential outcomes $Y_{i,t}(0)$ for treated units  are identifiable.


\vspace{-0.5cm}
\subsection{Notations}

For a constant $a \in \mathcal{R}$, denote $|a|$ as the absolute value of $a$. For a random variable $\mathbf{X}$, let $\Mean \left( \mathbf{X}\right)$ denote the expectation of $\mathbf{X}$.
For any vector $v$, denote $\|v\|_{0}$ as the number of non-zero entries of $v$.
For any matrix $\mathbf{M} \in \mathcal{R}^{m \times n}$, denote $\mathbf{M}^{\prime}$ as the transpose of matrix $\mathbf{M}$, and denote $\|\mathbf{M}\|_* =\sum_{i=1}^{\min \{m, n\}} \sigma_i(\mathbf{M}) $ as the nuclear norm.  Let $(\boldsymbol{U}, \boldsymbol{D}, \boldsymbol{V})=\operatorname{svd}( \mathbf{M})$ denote the singular value decomposition of $\mathbf{M}$, i.e.,  $\mathbf{M} =\boldsymbol{U}\boldsymbol{D} \boldsymbol{V}^{T}$, where $\boldsymbol{U} \in \mathcal{R}^{m \times \min\{m,n\}}$, $\boldsymbol{D}\in \mathcal{R}^{\min\{m,n\} \times \min\{m,n\}}$, $ \boldsymbol{V} \in \mathcal{R}^{n \times \min\{m,n\}}$, $\boldsymbol{U}^{\top} \boldsymbol{U}=\boldsymbol{V}^{\top} \boldsymbol{V}=\boldsymbol{I}_{\min\{m,n\}}$  and $\boldsymbol{D}_{k, k}=\sigma_k(\mathbf{M}) \geq 0$ for $k \in\{1, \ldots, s\}$. For any matrix $\mathbf{M}_1$ and  $\mathbf{M}_2$ with commensurate dimensions, let $\left <  \mathbf{M}_1, \mathbf{M}_2\right> = \operatorname{trace} \left(\mathbf{M}_2^T \mathbf{M}_1 \right) $ denote the trace inner product on matrix space. 
For any subspace $\mathcal{S}$, denote its orthogonal complement as $ \overline{x}^{\perp}:=\left\{\mathbf{N} \mid\langle \mathbf{M}, \mathbf{M} \rangle=0 \text { for all } \mathbf{M} \in \overline{\mathcal{S}}\right\}$.
For any matrix $\mathbf{M}$ and subspace $\mathcal{S}$, let $\mathbf{M}_\mathcal{S}$ be the components of $\mathbf{M}$ restricted to the support $\mathcal{S}$ , i.e. $\mathbf{M}_\mathcal{S} = argmin_{\mathbf{N} \in \mathcal{S}} \|\mathbf{M} - \mathbf{N} \|_{F} $ and similarly for $\mathbf{M}_{\overline{\mathcal{S}}^{\perp}}$.

\vspace{-0.5cm}
\section{Methodology}\label{sec:method}

In this section, we present our novel approach to efficiently estimate Synthetic Control  weights for multiple treated units. 

The classic Synthetic Control method \citep{abadie2010synthetic}  operates under the assumption that a weighted average of control units provides a good approximation for the counterfactual outcome of the treated unit as if it has been under control. Specifically, $\widehat{Y}_{i, T_0+1}(0)$ is estimated using
\vspace{-0.5cm}
\begin{equation}
\widehat{Y}_{i, T_0+1}(0)=\sum_{j=m+1}^{j=m+n} \widehat{\theta}_{i, j} Y_{j, T_0+1}
\end{equation}
for $i=1,2, \cdots m$, where $\widehat{\theta}_{i, j}$ 
 are determined using a constrained linear regression:
\begin{equation} \label{eq:SCforSingle}
\begin{aligned}
& \min \sum_{t=1}^{T_0}\left(Y_{i, t}-\sum_{j=m+1}^{j=m+n} \theta_{i, j} Y_{j, t}\right)^2,  \text{subject to } \left\{ \begin{array}{l}
\theta_{i, j} \geq 0 \\
\sum_{j=m+1}^{j=m+n} \theta_{i, j}=1 \text {, for } i=1,2, \cdots m.
\end{array} \right. 
\end{aligned}
\end{equation}
 For simplicity, we assume that no other predictors of the outcome are available and only regress on the outcome of control units. We notice that when the constraints are not applied, 
we are working on the following projection: 
\begin{equation*} 
\left(\begin{array}{c}
Y_{i,1}  \\
\vdots \\
Y_{i, T_0}
\end{array}\right)
=\left(\begin{array}{cccc}
Y_{m+1,1} & \ldots & Y_{m+n, 1} \\
\vdots & & \vdots \\
Y_{m+1, T_0} & \ldots & Y_{m+n, T_0}
\end{array}\right)
\left(\begin{array}{c}
\theta_{i,1}  \\
\vdots  \\
\theta_{i, n} 
\end{array}\right)+
\left(\begin{array}{c}
\varepsilon_{i,1}  \\
\vdots \\
 \varepsilon_{i, T_0}
\end{array}\right),
\end{equation*}
for each treated unit $i=1,2, \cdots m$. We note that the design matrices of the regression model for  each treated unit are exactly the same. This is due to the fact that we are regressing on the same set of control units for different treated units.

This allows us to represent the above Synthetic Control models for multiple treated units in a single matrix equation. Specifically, we express the relationship as:
\begin{equation} \label{eq:MatrixModel}
\mathbf{Y}_{T_0 \times m}=\mathbf{X}_{T_0 \times n} \mathbf{\Theta}_{n \times m}+\mathbf{E}_{T_0\times m},
\end{equation}
where   $\mathbf{Y}_{T_0 \times m}$ presents the pre-treatment information of the treated units, $\mathbf{X}_{T_0 \times n} $ denotes the pre-treatment information of the control units,  $\mathbf{\Theta}_{n \times m} $ denotes the coefficient matrix and $\mathbf{E}_{T_0\times m}$ is the error matrix:
\begin{equation*}
\mathbf{Y}_{T_0 \times m} = \left(\begin{array}{ccc}
Y_{1,1} & \ldots & Y_{m, 1} \\
\vdots & & \vdots \\
Y_{1, T_0} & \ldots & Y_{m, T_0}
\end{array}\right), \mathbf{X}_{T_0 \times n}  = \left(\begin{array}{cccc}
Y_{m+1,1} & \ldots & Y_{m+n, 1} \\
\vdots & & \vdots \\
Y_{m+1, T_0} & \ldots & Y_{m+n, T_0}
\end{array}\right)
\end{equation*}
\begin{equation*}
\mathbf{\Theta}_{n \times m} = \left(\begin{array}{cccccc}
\theta_{1,1} & \ldots & \theta_{m, 1} \\
\vdots & & \vdots \\
\theta_{1, n} & \ldots & \theta_{m, n}
\end{array}\right), \mathbf{E}_{T_0\times m} = \left(\begin{array}{ccc}
\varepsilon_{1,1} & \ldots & \varepsilon_{m, 1} \\
\vdots & \vdots & \vdots \\
 \varepsilon_{1, T_0} & \ldots & \varepsilon_{m, T_0}
\end{array}\right)
\end{equation*}

 In our setting, we assume that the coefficient matrix $\mathbf{\Theta}_{n \times m}$ is independent of time $t$.

For cases where $T_0 > n$, the above equation is a classic Multivariate Linear Regression problem, with an Ordinary Least Squares (OLS) estimator, $\widehat{\mathbf{\Theta}}_{n \times m}^{O L S}=\left(\mathbf{X}^{\prime}\mathbf{X}\right)^{-1} \mathbf{X}^{\prime} \mathbf{Y}.$ The computational complexity can be reduced to $O\left(n^3+T_0n^2+T_0mn \right)$.
Compared to fitting separate SC models for each unit treated in Equation \eqref{eq:SCforSingle} with order $O\left(m n^2\left(T_0+n\right)\right)$, we can significantly save computational time since we do not need to compute the projection matrix $\left(\mathbf{X}^{\prime}\mathbf{X}\right)^{-1} \mathbf{X}^{\prime} $ repeatedly.

However, in high-dimensional scenarios with $n > T_0$, the OLS estimator does not exist. We address this challenge by adopting the Multivariate Square-root Lasso method, expressed as:
\vspace{-0.5cm}
\begin{equation} \label{eq:Object}
\underset{\mathbf{\Theta} \in \mathbb{R}^{n \times m}}{\min }\left\{ \mathcal{\mathcal{L}(\mathbf{\Theta})}: = \frac{1}{\sqrt{T_0}}\|\mathbf{Y}-\mathbf{X}  \mathbf{\Theta}\|_* + \lambda \sum_{i=1}^{n} \sum_{j=1}^{m} |\mathbf{\Theta}_{i,j}|\right\}.
\end{equation}
We choose the Multivariate Square-root Lasso  for several reasons. 
Firstly, we are dealing with a high-dimensional setting and the Lasso regularizer is well-known for conducting dimension reduction and coefficient estimation simultaneously in linear models \citep{tibshirani1996regression,lounici2008sup, obozinski2011support}. 
Secondly, the Square-Root Lasso \citep{belloni2011square,sun2012scaled} was proven to be pivotal  such that  the selection of   tuning parameter  does not depend on the unknown variance estimator.
Thirdly, the  Multivariate Square-root Lasso implicitly estimates the error covariance and performs similarly to methods that explicitly estimate the error covariance in terms of Frobenius norm error \citep{molstad2021new}.
Notably, it is characterized by its computational efficiency and convexity, making it a practical and reliable choice for our purposes.

 In the rest of the paper, we denote the solution for the problem \eqref{eq:Object} as
 \vspace{-0.3cm}
\begin{equation} \label{eq:estimator}
	\widehat{\mathbf{\Theta}} = \underset{\mathbf{\Theta} \in \mathbb{R}^{n \times m}}{\arg \min }\left\{ \mathcal{\mathcal{L}(\mathbf{\Theta})}\right\}.
\end{equation}
And denote $\mathbf{\Theta}^{*}$ as the optimal solution, that is,
 \begin{equation}
 	\mathbf{\Theta}^{*}=\underset{\mathbf{\Theta} \in \mathcal{R}_{n \times m}}{\arg \min }\left\{ \frac{1}{\sqrt{T_0}}\|\mathbf{Y}-\mathbf{X}  \mathbf{\Theta}\|_* \right\}.
 \end{equation}
 
Given the estimated $\widehat{\mathbf{\Theta}} $ in hand, we are able to estimate $\widehat{\mathbf{Y}}_{T_0 \times m}=\mathbf{X}_{T_0 \times n} \widehat{\mathbf{\Theta}}_{n \times m}$, and then estimate the ATT  as follows $\widehat{\delta}   =  \frac{1}{m}\sum_{i=1}^{m} \left\{ Y_{i,T_0+1} - \widehat{Y}_{i,T_0+1}(0) \right\} .$x

\section{Statistical Guarantees}\label{sec:theory}

In this section,  we demonstrate the validity of our proposed method by deriving the estimation error bounds.  Our goal is to provide finite sample bounds for two key aspects: the Frobenius norm of the difference between the estimated coefficient matrix and its true value , denoted as $\widehat{\mathbf{\Theta}}-\mathbf{\Theta}^*$, and bias of ATT estimation,  denoted as $\widehat{\delta} - \delta$. To establish the theoretical guarantee, we will require the following assumptions.


\begin{assumption} \label{ass:sparse}
The coefficient matrix $\mathbf{\Theta}^{\star}$ is s-sparse, with $\left|\operatorname{vec}\left(\mathbf{\Theta}^{\star}\right)\right|_0 = s$.
\end{assumption}


%
Assumption \ref{ass:sparse} is standard in high-dimensional literature and requirements for the sparsity parameter $s$ will be discussed after Corollary \ref{coro}. In the following, we denote the  the support of $\mathbf{\Theta}^{*}$ as $\mathcal{S}$.  In addition, following \cite{molstad2021new},  for $g(\mathbf{\Theta})=  \sum_{i=1}^{n} \sum_{j=1}^{m} |\mathbf{\Theta}_{i,j}|$,  and any constant $c>1$, we introduce a quantity $\phi_{\mathbf{E}, g}(\mathcal{S}, c)$  defined as:
\begin{equation}
\inf _{\boldsymbol{\Delta} \in \mathcal{C}_g(\mathcal{S}, c)}\left\{\frac{\sup _{\|\boldsymbol{Q}\|_{*} \leq 1} \operatorname{tr}\left\{\left(\boldsymbol{Q}-\boldsymbol{U}_\epsilon \boldsymbol{V}_\epsilon^{\top}\right)^{\top}(\mathbf{E}-\boldsymbol{X} \Delta)\right\}}{\sqrt{T_0}\|\boldsymbol{\Delta}\|_F^2}\right\}, 
\end{equation}
with $ \mathcal{C}_g(\mathcal{S}, c)=\left\{\boldsymbol{\Delta} \in \mathbb{R}^{n \times m}: \boldsymbol{\Delta} \neq 0, g\left(\boldsymbol{\Delta}_{\overline{\mathcal{S}}^{\perp}}\right) \leq \frac{c+1}{c-1} g\left(\boldsymbol{\Delta}_{\overline{\mathcal{S}}}\right)\right\}$, 
where $\quad\left(\boldsymbol{U}_{\mathbf{E}}, \boldsymbol{D}_{\mathbf{E}},\boldsymbol{V}_{\mathbf{E}}\right)=\operatorname{svd}(\mathbf{E})$.  And we impose the following technical assumption:
\begin{assumption} \label{ass:RSC}
There exists a constant $b$ such that $\phi_{\mathbf{E}, g}(\mathcal{S}, c) \geq b>0$ almost surely. 
\end{assumption}

Assumption \ref{ass:RSC} is closely related to the restricted strong convexity \citep{negahban2012unified} of the nuclear norm of the error matrix $\mathbf{E}$, but also depends on $\mathbf{X}$. 
Under these assumptions, we present a theorem that establishes an estimation error bound on the Frobenius norm of the difference between the estimated coefficient matrix and its true value $\widehat{\mathbf{\Theta}}-\mathbf{\Theta}^*$ as follows.

\begin{thm}{(Estimation Error)}  \label{thm:ErrorBound}
Assume Assumption  \ref{ass:noAnticipation}, \ref{ass:Consistency},  \ref{ass:sparse}, and \ref{ass:RSC} hold. For any fixed constant $c>1$, and $\lambda \geq \frac{c}{\sqrt{T_0}}  \left\{  \tilde{g}\left(\mathbf{X}^{\top} \mathbf{U}_{\mathbf{E}} \mathbf{V}_{\mathbf{E}}^{\top}\right) + sup_{ \mathbf{Z} \in \Lambda} \tilde{g}\left(\mathbf{X}^{\top} \mathbf{Z}\right)\right\}$, with
\begin{equation*}
\Lambda: =\left\{ \mathbf{Z}:\mathbf{Z}\in \mathcal{R}^{T_0\times m}, \|\mathbf{Z}\|_2 \leq 1, \mathbf{U_{\mathbf{E}}}^{\top} \mathbf{Z}=0, \mathbf{Z} \mathbf{V}_{\mathbf{E}}=0\right\},
\end{equation*}
the estimation in Equation \eqref{eq:Object} satisfies
\begin{equation*} 
	\|\widehat{\mathbf{\Theta}}-\mathbf{\Theta}^*\|_F \leq  \frac{(c+1)\lambda \sqrt{s}}{c \phi_{\mathbf{E}, g}(\mathcal{S}, c)} .
\end{equation*}
\end{thm}
This theorem provides a quantitative evaluation of the estimation accuracy under mild conditions, shedding light on the validity of our proposed method within the Synthetic Control framework.  The estimation error is directly proportional to both $\lambda$ and the square root of $s$. This indicates that $s$ increases, the problem becomes more challenging to solve. Additionally, larger values of 
$\lambda$ could potentially result in greater estimation errors.  Theorem \ref{thm:ErrorBound} also suggests that the selection of $\lambda$ is determined solely by $\mathbf{X}$, $\mathbf{U_{\mathbf{E}}}$, and $\mathbf{V}_{\mathbf{E}}$, without any dependence on the unknown covariance structure of $\mathbf{E}$.  
With two additional assumptions on the distribution of the potential outcomes, we further derive a simplified error bound in Corollary \ref{coro}.
 
\begin{assumption}\label{ass:BoundedMean}
The mean of the potential outcome $Y_{i, t}(0)$ at time $t$, i.e.,  $\Mean \left\{ Y_{i, t}(0) \right\}$ is bounded by $L >0$.
\end{assumption}
\begin{assumption}\label{ass:subGaussian}
The potential outcome $Y_{i, t}(0)$ is $\sigma$-sub Gaussian, i.e., $\mathbb{E}(\exp \{c Y_{i, t}(0)\}) \leq \exp \left\{c^2 \sigma^2 / 2\right\}$ for all $c \in \mathbb{R}$, $i=1,2,\cdots,N$, and $t=1,2,\cdots, T$.
\end{assumption}

Assumption \ref{ass:subGaussian} requires the tail performance of the potential outcome $Y_{i, t}(0)$ and does not exclude the possibility of dependency among the potential outcome $Y_{i, t}(0)$ at various time points for various units.  
\begin{coro} \label{coro}
Assume Theorem \ref{thm:ErrorBound} with Assumption \ref{ass:BoundedMean} and \ref{ass:subGaussian}  hold, then for \\ $\lambda \geq  2c\left\{n \log(nT_0)/T_0 \right\}^{1/4}$,
the estimation in Equation \eqref{eq:Object} satisfies
\begin{equation} \label{eq:ErrorBound}
	\|\widehat{\mathbf{\Theta}}-\mathbf{\Theta}^*\|_F \leq  \frac{(c+1) \lambda \sqrt{s}  }{c \phi_{\mathbf{E}, g}(\mathcal{S}, c)} ,
\end{equation}
with probability greater than $1-\sqrt{2} \sigma\left\{ \log(nT_0)/(nT_0) \right\}^{1/4}.$
\end{coro}
\vspace{-0.5em}
Corollary \ref{coro} serves as a valuable tool for investigating the performance of our method in various scenarios of sparsity.  In the case of 'hard'-sparsity, where the degree of sparsity remains relatively constant as the dataset scales (i.e., $s$ is a constant with respect to $n$), Corollary \ref{coro} reveals that, by setting $\lambda = 2c\left\{n \log(nT_0)/T_0 \right\}^{1/4}$, we can guarantee the consistency of our coefficient matrix estimator. Specifically, the estimation error satisfies
the order of $O_p\left(  \left\{\log(nT_0)/(nT_0) \right\}^{1/4}\right)$,
which goes to zero as $nT_0$ goes to infinity.

In the case of 'soft'-sparsity, when  $\lambda = 2c\left\{n \log(nT_0)/T_0 \right\}^{1/4}$, and the degree of sparsity satisfies $s =O_p\left( \left\{n/log(n)\right\}^{1/4}\right)$, the estimation error then satisfies the order of $O_p\left( \left\{\log(nT_0)/(nT_0)\right\}^{1/8} \right)$,
 which goes to zero as $nT_0$ goes to infinity and implies the consistency of our proposed coefficient matrix estimator in terms of Frobenius norm.

We remark that our estimator and its theoretical analysis are motivated by and generalize existing research on  Multivariate Square-root Lasso \citep{van2016chi,molstad2021new}.  However, our established estimation rate offers a novel contribution to the field by addressing unique challenges encountered within the Synthetic Control framework.

 Firstly, in the context of disaggregated data, we face a substantial issue of high-dimensionality, in the sense that there are more unknown parameters (units within the synthetic control framework) than available data points (time points in the synthetic control framework). To the best of our knowledge, existing literature on Multivariate Square-root Lasso has not fully tackled this high-dimensional challenge. We are the first to derive an error bound in such a high-dimensional setting.

Secondly,  previous literature \citep{van2016chi,molstad2021new} considers fixed design matrices, which is not suitable within the Synthetic Control framework. In contrast, in Theorem \ref{thm:ErrorBound} and Corollary \ref{coro}, we consider a random design matrix.

Lastly, the outcome of interest in the synthetic control framework exhibits significant temporal correlations, making estimation more challenging. Previous work on Multivariate Square-root Lasso \citep{molstad2021new} requires that the distribution of error matrix $\mathbf{E}$ to be left-spherical, which might not always hold true, especially in cases with time-dependent data.  In contrast, our theoretical results  imposes no specific assumptions concerning the covariance structure of the design matrix $\mathbf{X}$ or the error matrix $\mathbf{E}$. This flexibility allows our approach to be applied to panel data with ease.

With Corollary \ref{coro} in hand, we further investigate the estimation bias of ATT, namely,  $\widehat{\delta} - \delta$ under the same condition.

\begin{thm}{(ATT Estimation Error)}  \label{thm:ATT}
Assume that Corollary \ref{coro} holds, then the estimated Average Treatment Effect $\widehat{\delta}$ enjoys the learning rate:
\begin{equation*}
\widehat{\delta} - \delta  \leq   \left\{ \frac{T_0}{log(T_0)}\right\} ^{1/8}\frac{(c+1) \sqrt{ns} \lambda }{c \phi_{\mathbf{E}, g}(\mathcal{S}, c)\sqrt{m}},
 \end{equation*}
with probability greater than  $1-\sqrt{2} \sigma\left\{ \log(nT_0)/(nT_0) \right\}^{1/4} -  \left( \sigma^2+L^2\right) \left\{ \frac{log(T_0)}{T_0}\right\} ^{1/8}.$
\end{thm}

Theorem \ref{thm:ATT} reveals that when the number of control units $n$ is fixed, the upper bound goes to zero as m goes to infinity, assuming a fixed 'hard'-sparsity level s.  While in the case of 'soft'-sparsity, the upper bound goes to zero as long as $\sqrt{s/m} \rightarrow 0$ as $m \rightarrow$ 0.  

%
%
%

\section{Simulation Studies}\label{sec:simulation}

In this section, we illustrate the validity and effectiveness of our proposed method via extensive simulation studies. Specifically, we consider the scenario with $T_0=100$ as the pre-treatment time periods, $T_1=10$ as the post-treatment time periods, $m=$ 50,150,200, 250, 300,350,400 treated units,  and $n=400$ control units.  We consider generating  $\mathbf{X}_{T_0 \times n} $ using an AR(1) model $Y_{i, t}(0)=0.1*c_i+0.9Y_{i, t-1}(0)+Z_{i, t}$ with $c_i \in\{1,2,\cdots, 10\}$ 
and $Z_{i, t} \sim N(0,1)$.

%

We consider two ways to generate $\mathbf{Y}$ for treated units: \textbf{Setting (1):} We generate $\mathbf{Y}$ using the same procedure as we did for  $\mathbf{X}_{T_0 \times n} $ . \textbf{Setting (2):} We generate $\mathbf{Y}_{T_0 \times m}$ using Equation \eqref{eq:MatrixModel}. Here, $\mathbf{\Theta}$ is a random matrix with $s=1000$, with each column summing to 1, and $\mathbf{E}_{t,i} \stackrel{i . i . d}{\sim} N\left(0,0.5^2\right)$.

We compare the performance of our proposed method, denoted as MSC for brevity,  with three baselines for estimating  $Y_{i, t}(0)$ after the treatment assignment. These baselines include: (1) PSC: fitting penalized SC method separately for each treated unit \citep{abadie2021penalized}; (2)SCUL: fitting Synthetic Control Using Lasso  separately for each treated unit \citep{hollingsworth2020tactics}; (3)ROLS: fitting  restricted OLS separately for each treated unit \citep{chernozhukov2021exact}. For MSC, PSC, and SCUL, the penalty parameters are predetermined through cross-validation. Additionally, we use a hyperparameter value of 1 for ROLS, as recommended by \cite{chernozhukov2021exact}. A sensitivity analysis of the parameter tuning for MSC is available in  Appendix \ref{sec:additionalSimu}. All evaluations in this simulation study are based on a single run of each method using the pre-selected hyperparameters.

\begin{figure}[!b]
\centering
\includegraphics[width=  0.75\linewidth]{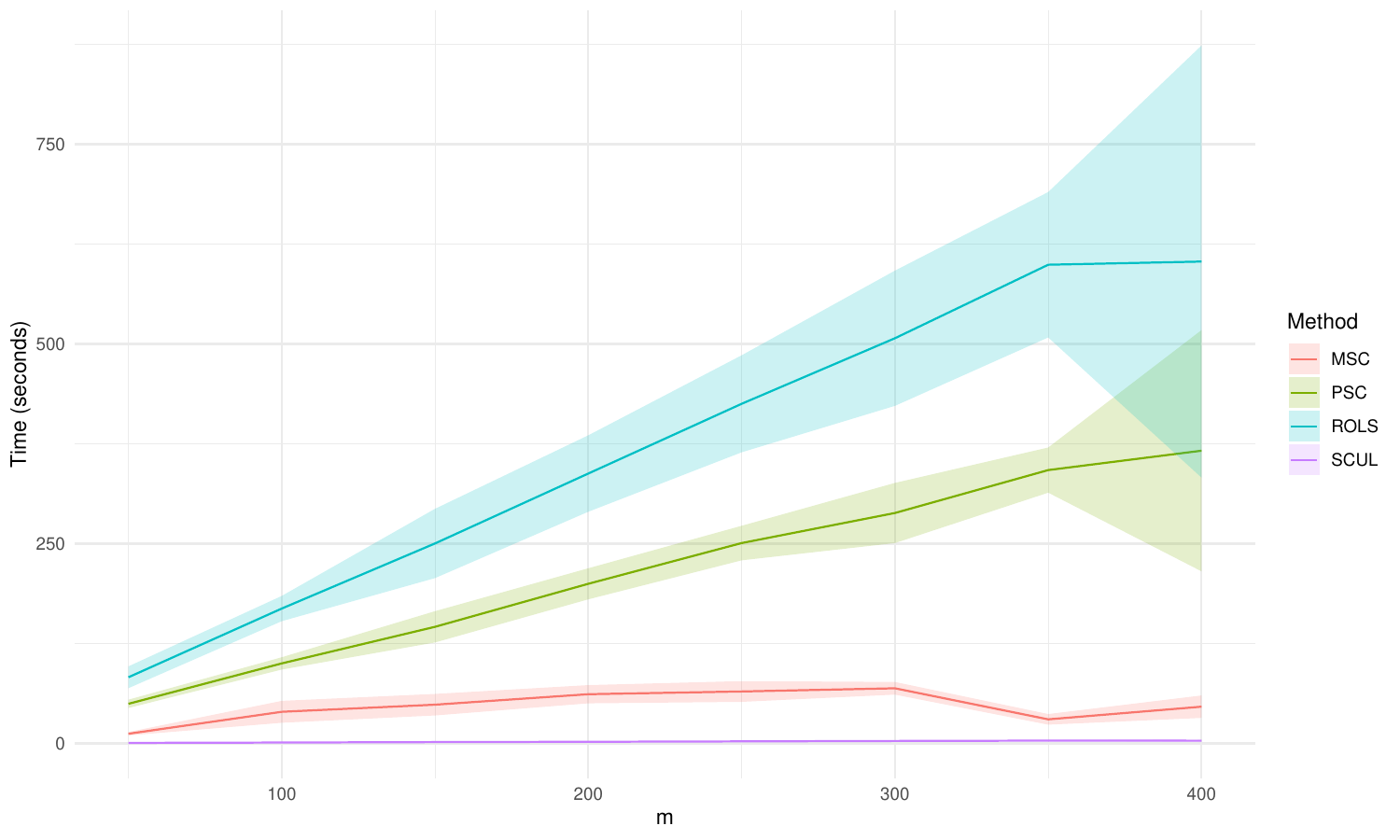}
\caption{Computational time analysis for various methods under Setting (2)   with $T_0=100$ pre-treatment periods, $T_1=10$ post-treatment periods, $m=50,150,200,250,300,350,400$ treated units, $n=400$ control units and $s=1000$. The experiments are repeated 500 times, with the solid line representing the average and the shadow area representing one standard error.}
\label{fig:Time2}
\vspace{-1 em}
\end{figure}

\begin{figure}[!h]
\centering
\includegraphics[width=0.42\linewidth]{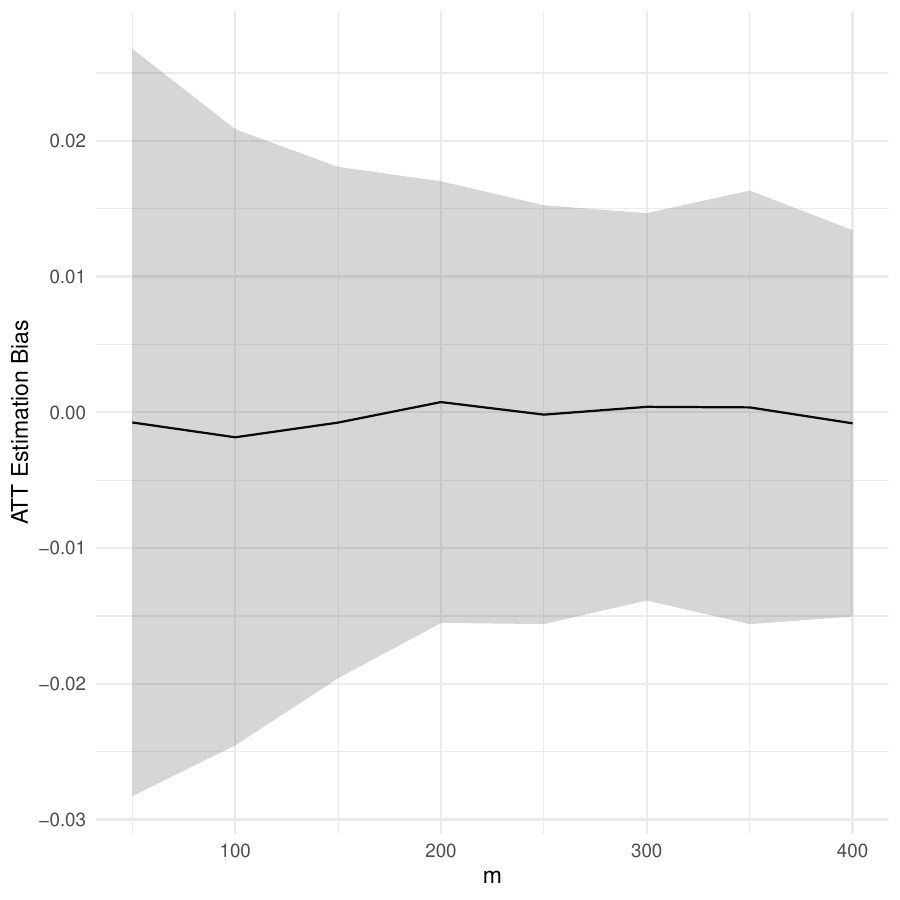}
\caption{ATT estimation bias for MSC under Setting (2)   with $T_0=100$ pre-treatment periods, $T_1=10$ post-treatment periods, $m=50,150,200,250,300,350,400$ treated units, $n=400$ control units and $s=1000$. The experiments are repeated 500 times in total, with the solid line representing the average and the shadow area representing one standard error.}
\label{fig:ATT2_trend}
\end{figure}

\begin{figure*}[!h]
\centering
\includegraphics[width=\textwidth]{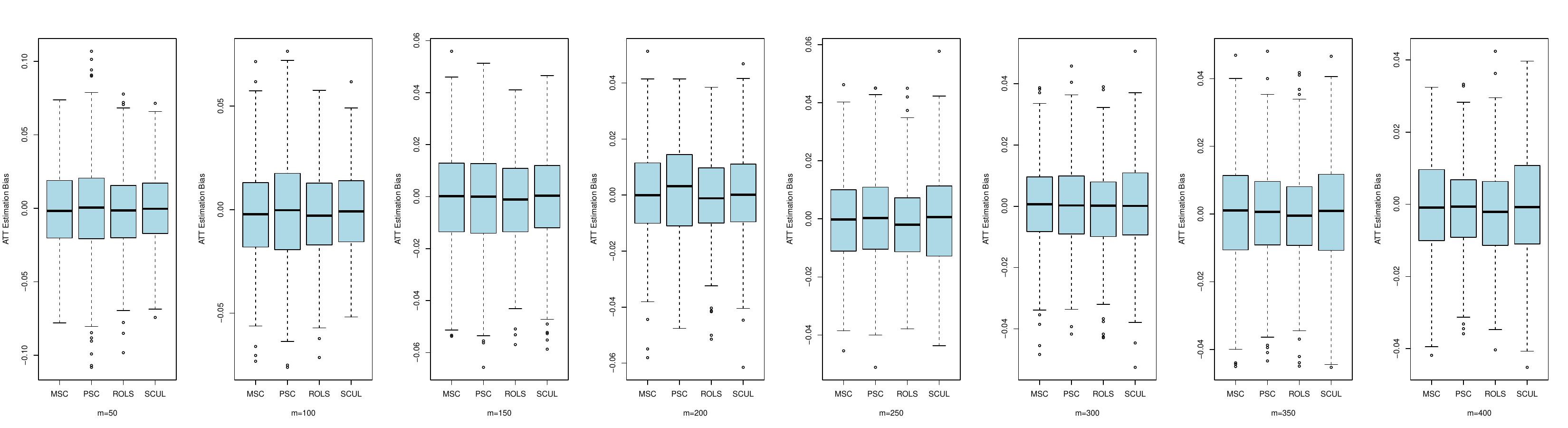}
\caption{ATT Estimation Bias of independent 500 runs under Setting  (2)    with $T_0=100$ pre-treatment periods, $T_1=10$ post-treatment periods,  $m=$ 50,150, 200, 250, 300, 350, 400 treated units, $n=400$ control units and $s=1000$.} 
\label{fig:att2}
\end{figure*}

We employ three key metrics to assess the performance of these methods:
\begin{itemize}	
\item Computational time: we compare the time it takes to compute the Synthetic Control weights for a single run of each methods without cross-validation in Figure \ref{fig:Time2};
\item  ATT Estimation Bias: we present box plots of the ATT estimation bias $\widehat{\delta} - \delta$ for each method in Figure \ref{fig:ATT2_trend} and Figure \ref{fig:att2};
	\item 	Root Mean Squared Error (RMSE) for estimating  $Y_{i, t}(0)$ after $T_0$:
    \begin{equation*}
R M S E= \sqrt{\frac{1}{m T_1} \sum_{t=T_0+1}^{T_0+T_1} \sum_{i=1}^m\left(\widehat{Y}_{i, t}(0)-Y_{i, t(0)}\right)^2};
\end{equation*}
We summarize the mean of MSEs in Table \ref{tb:mse2}.
\end{itemize}
with additional simulation results   and  in Appendix \ref{sec:additionalSimu}.

\begin{table*}[h]
\centering
\caption{The corresponding RMSE means of independent 500 runs for fitting Synthetic Control weights under Setting  (2) with $T_0=100$ pre-treatment periods, $T_1=10$ post-treatment periods, $m=50,150,200,250,300,350,400$ treated units, $n=400$ control units and $s=1000$.}	
\label{tb:mse2}
\begin{tabular}{|r|r|r|r|r|r|r|r|r|}
  \hline
 m & 50 & 100 & 150 & 200 & 250 & 300 & 350 & 400 \\ 
  \hline
MSC & 0.71 & 0.71 & 0.72 & 0.72 & 0.72 & 0.72 & 0.73 & 0.73 \\    \hline
  PSC & 0.72 & 0.72 & 0.72 & 0.73 & 0.73 & 0.73 & 0.73 & 0.73 \\    \hline
  ROLS & 0.72 & 0.72 & 0.72 & 0.72 & 0.72 & 0.73 & 0.73 & 0.73 \\    \hline
  SCUL & 1.04 & 1.22 & 1.34 & 1.45 & 1.53 & 1.61 & 1.68 & 1.73 \\ \hline
\end{tabular}
\end{table*}

It is clear from Figure \ref{fig:Time2} that when $m$ is large, the computational time of our proposed MSC is significantly less than that of PSL and ROLS. Remarkably, this efficiency gain does not come at the cost of increased  RMSE as demonstrated in  Table \ref{tb:mse2}. 
It is also noteworthy that SCUL is computationally efficient due to the efficient performance of R function 'glmnet'. However, it suffers from high RMSE. 

For the ATT estimation bias, as illustrated in Figure \ref{fig:ATT2_trend} and Figure \ref{fig:att2}, our proposed MSC consistently exhibits unbiased behavior. Furthermore, the variance decreases as the number of treated units $m$ increases, with levels comparable to the baseline methods. In summary, our MSC method not only significantly reduces computation time but also maintains estimation accuracy.


\section{Real Data Application}\label{sec:application}

After the COVID-19 outbreak in 2020, most states imposed Stay-at-Home Orders. However, seven states - Arkansas, Iowa, Nebraska, North Dakota, South Dakota, Utah, and Wyoming-chose not to implement these orders, which are marked in red on the map in Figure \ref{fig:map}. These Stay-at-Home Orders were mostly put into effect in late March or early April, as detailed in Table \ref{tb:SAHtime} in the Appendix.
Given that these Stay-at-Home orders were implemented in partial states around the same time, it is an opportunity to measure their impact on unemployment rates using economic methods like Synthetic Control and Difference in Difference\citep{gibson2020understanding, beland2020covid,baek2021unemployment}.


We apply our method to investigate the causal effect of COVID-19 Stay-at-Home Orders on the monthly unemployment rate in conterminous United States at the county level. 
Our dataset includes monthly unemployment rate data from January 2010 to April 2020, obtained from the official website of the Bureau of Labor Statistics' Local Area Unemployment Statistics (LAUS) program conducted by the Bureau of Labor Statistics (BLS) \footnote{Official website of the Local Area Unemployment Statistics  program: \url{https://www.bls.gov/lau/#data}}, same as \cite{baek2021unemployment}.  
The BLS primarily relies on the  Current Population Survey (CPS) for constructing county-level employment and unemployment estimates. 
Fortunately, the survey reference week for the CPS for March 2020 was March 8 through March 14, and the reference week for April was April 12 through April 18 \footnote{For further details on the methodology used by the Bureau of Labor Statistics, please visit \url{https://www.bls.gov/lau/laumthd.htm}.} \citep{baek2021unemployment}, which  align quite nicely with the broad implementation of  Stay-at-Home Orders.

In our dataset, we have a total of 3,112 counties, with $n=438$  control counties and $m=2674$ treated counties. We consider April 2020 as the beginning of the treatment period, with one month post-treatment and $T_0=147$ months pre-treatment. 
We plot the trend of mean unemployment rate in Figure \ref{fig:realdatamean} with a county-level spaghetti plot in Figure \ref{fig:spaghetti}. We noticed a rapid rise in the unemployment rate in April 2020.
Upon applying our proposed method, we found that the implementation of COVID-19 Stay-at-Home Orders led to a notable increase in the monthly unemployment rate in the contiguous United States at the county level. Specifically, we observed an average of 5.06 percentage point rise
of the unemployment rate in counties that employ the Stay-at-Home Orders. This outcome aligns closely with the findings of \cite{baek2021unemployment} , who reported a 1.5 (SE: 0.331) percentage point increase in the unemployment rate each week.

\section{Discussion}\label{sec:discussion}

In this paper, we propose an innovative approach that leverages the Multivariate Square-root Lasso to fit Synthetic Control weights for multiple treated units. Our method exhibits a remarkable reduction in computation time while maintaining estimation accuracy, as supported by both theoretical analysis and numerical experiments.  Different from learning weights for each treated unit iteratively, the weights learned by our approach emphasize the sparsity of the whole coefficient matrix rather than the sparsity of the weights for individual treated unit.
However, our proposed method requires that treatment assignment time is the same for all treated units. This assumption restricts the applicability of MSC when treatments are staggered or have varying start dates across units. 
Looking ahead, there are several exciting avenues to explore for future work.
Firstly, we could explore extensions or adaptations of MSC to accommodate staggered treatment adoption.
Secondly, extending the Square-Root Lasso to incorporate more advanced penalty terms, as explored by \cite{abadie2021penalized}, could further enhance the method's flexibility and performance. 
In addition, integrating auxiliary information into the model fitting process, when available, can lead to improved model performance.


\bibliography{mycite}

@article{splawa1990application,
	author = {Splawa-Neyman, Jerzy and Dabrowska, Dorota M and Speed, TP},
	journal = {Statistical Science},
	pages = {465--472},
	publisher = {JSTOR},
	title = {On the application of probability theory to agricultural experiments. Essay on principles. Section 9.},
	year = {1990}}

@article{rubin1974estimating,
	author = {Rubin, Donald B},
	journal = {Journal of educational Psychology},
	number = {5},
	pages = {688},
	publisher = {American Psychological Association},
	title = {Estimating causal effects of treatments in randomized and nonrandomized studies.},
	volume = {66},
	year = {1974}}

@article{baek2021unemployment,
  title={Unemployment effects of stay-at-home orders: Evidence from high-frequency claims data},
  author={Baek, ChaeWon and McCrory, Peter B and Messer, Todd and Mui, Preston},
  journal={Review of Economics and Statistics},
  volume={103},
  number={5},
  pages={979--993},
  year={2021},
  publisher={MIT Press One Rogers Street, Cambridge, MA 02142-1209, USA 
}
}

@article{athey2016recursive,
  title={Recursive partitioning for heterogeneous causal effects},
  author={Athey, Susan and Imbens, Guido},
  journal={Proceedings of the National Academy of Sciences},
  volume={113},
  number={27},
  pages={7353--7360},
  year={2016},
  publisher={National Acad Sciences}
}

@article{abadie2021using,
	author = {Abadie, Alberto},
	journal = {Journal of Economic Literature},
	number = {2},
	pages = {391--425},
	title = {Using synthetic controls: Feasibility, data requirements, and methodological aspects},
	volume = {59},
	year = {2021}}

@article{beland2020covid,
  title={COVID-19, stay-at-home orders and employment: Evidence from CPS data},
  author={Beland, Louis-Philippe and Brodeur, Abel and Wright, Taylor},
  year={2020},
  publisher={IZA Discussion Paper}
}

@article{turlach2005simultaneous,
	author = {Turlach, Berwin A and Venables, William N and Wright, Stephen J},
	journal = {Technometrics},
	number = {3},
	pages = {349--363},
	publisher = {Taylor \& Francis},
	title = {Simultaneous variable selection},
	volume = {47},
	year = {2005}}

@article{yuan2007dimension,
	author = {Yuan, Ming and Ekici, Ali and Lu, Zhaosong and Monteiro, Renato},
	journal = {Journal of the Royal Statistical Society: Series B (Statistical Methodology)},
	number = {3},
	pages = {329--346},
	publisher = {Wiley Online Library},
	title = {Dimension reduction and coefficient estimation in multivariate linear regression},
	volume = {69},
	year = {2007}}

@article{obozinski2011support,
	author = {Obozinski, Guillaume and Wainwright, Martin J and Jordan, Michael I},
	journal = {The Annals of Statistics},
	number = {1},
	pages = {1--47},
	publisher = {Institute of Mathematical Statistics},
	title = {Support union recovery in high-dimensional multivariate regression},
	volume = {39},
	year = {2011}}

@article{negahban2011estimation,
	author = {Negahban, Sahand and Wainwright, Martin J},
	journal = {The Annals of Statistics},
	number = {2},
	pages = {1069--1097},
	publisher = {Institute of Mathematical Statistics},
	title = {Estimation of (near) low-rank matrices with noise and high-dimensional scaling},
	volume = {39},
	year = {2011}}

@article{molstad2021new,
	author = {Molstad, Aaron J},
	title = {New insights for the multivariate square-root lasso},
	year = {2021}}

@article{rothman2010sparse,
	author = {Rothman, Adam J and Levina, Elizaveta and Zhu, Ji},
	journal = {Journal of Computational and Graphical Statistics},
	number = {4},
	pages = {947--962},
	publisher = {Taylor \& Francis},
	title = {Sparse multivariate regression with covariance estimation},
	volume = {19},
	year = {2010}}

@inproceedings{van2016chi,
  title={$\chi$ 2-confidence sets in high-dimensional regression},
  author={van de Geer, Sara and Stucky, Benjamin},
  booktitle={Statistical Analysis for High-Dimensional Data: The Abel Symposium 2014},
  pages={279--306},
  year={2016},
  organization={Springer}
}

@article{niu2019simultaneous,
	author = {Niu, Xiaomeng and Cho, Hyunkeun Ryan},
	journal = {Communications in Statistics-Theory and Methods},
	number = {11},
	pages = {2734--2747},
	publisher = {Taylor \& Francis},
	title = {Simultaneous estimation and inference for multiple response variables},
	volume = {48},
	year = {2019}}

@article{chang2022robust,
	author = {Chang, Le and Welsh, AH},
	journal = {Journal of Computational and Graphical Statistics},
	pages = {1--13},
	publisher = {Taylor \& Francis},
	title = {Robust multivariate lasso regression with covariance estimation},
	year = {2022}}

@article{molstad2021explicit,
	author = {Molstad, Aaron J and Weng, Guangwei and Doss, Charles R and Rothman, Adam J},
	journal = {Journal of Computational and Graphical Statistics},
	number = {3},
	pages = {612--621},
	publisher = {Taylor \& Francis},
	title = {An explicit mean-covariance parameterization for multivariate response linear regression},
	volume = {30},
	year = {2021}}

@article{negahban2012unified,
  title={A unified framework for high-dimensional analysis of $ M $-estimators with decomposable regularizers},
  author={Negahban, Sahand N and Ravikumar, Pradeep and Wainwright, Martin J and Yu, Bin},
  journal={Statistical science},
  volume={27},
  number={4},
  pages={538--557},
  year={2012},
  publisher={Institute of Mathematical Statistics}
}

@article{tibshirani1996regression,
  title={Regression shrinkage and selection via the lasso},
  author={Tibshirani, Robert},
  journal={Journal of the Royal Statistical Society: Series B (Methodological)},
  volume={58},
  number={1},
  pages={267--288},
  year={1996},
  publisher={Wiley Online Library}
}

@article{belloni2011square,
  title={Square-root lasso: pivotal recovery of sparse signals via conic programming},
  author={Belloni, Alexandre and Chernozhukov, Victor and Wang, Lie},
  journal={Biometrika},
  volume={98},
  number={4},
  pages={791--806},
  year={2011},
  publisher={Oxford University Press}
}

@article{sun2012scaled,
  title={Scaled sparse linear regression},
  author={Sun, Tingni and Zhang, Cun-Hui},
  journal={Biometrika},
  volume={99},
  number={4},
  pages={879--898},
  year={2012},
  publisher={Oxford University Press}
}

@article{abadie2003economic,
  title={The economic costs of conflict: A case study of the Basque Country},
  author={Abadie, Alberto and Gardeazabal, Javier},
  journal={American economic review},
  volume={93},
  number={1},
  pages={113--132},
  year={2003}
}

@article{abadie2010synthetic,
	author = {Abadie, Alberto and Diamond, Alexis and Hainmueller, Jens},
	journal = {Journal of the American statistical Association},
	number = {490},
	pages = {493--505},
	publisher = {Taylor \& Francis},
	title = {Synthetic control methods for comparative case studies: Estimating the effect of California's tobacco control program},
	volume = {105},
	year = {2010}}

@article{abadie2021penalized,
  title={A penalized synthetic control estimator for disaggregated data},
  author={Abadie, Alberto and L’Hour, J{\'e}r{\'e}my},
  journal={Journal of the American Statistical Association},
  number={just-accepted},
  pages={1--34},
  year={2021},
  publisher={Taylor \& Francis}
}

@article{dube2015pooling,
  title={Pooling multiple case studies using synthetic controls: An application to minimum wage policies},
  author={Dube, Arindrajit and Zipperer, Ben},
  year={2015},
  publisher={IZA Discussion Paper}
}

@article{robbins2017framework,
  title={A framework for synthetic control methods with high-dimensional, micro-level data: evaluating a neighborhood-specific crime intervention},
  author={Robbins, Michael W and Saunders, Jessica and Kilmer, Beau},
  journal={Journal of the American Statistical Association},
  volume={112},
  number={517},
  pages={109--126},
  year={2017},
  publisher={Taylor \& Francis}
}

@article{sabia2012effects,
  title={Are the effects of minimum wage increases always small? New evidence from a case study of New York state},
  author={Sabia, Joseph J and Burkhauser, Richard V and Hansen, Benjamin},
  journal={Ilr Review},
  volume={65},
  number={2},
  pages={350--376},
  year={2012},
  publisher={SAGE Publications Sage CA: Los Angeles, CA}
}

@techreport{doudchenko2016balancing,
  title={Balancing, regression, difference-in-differences and synthetic control methods: A synthesis},
  author={Doudchenko, Nikolay and Imbens, Guido W},
  year={2016},
  institution={National Bureau of Economic Research}
}

@article{acemoglu2016value,
  title={The value of connections in turbulent times: Evidence from the United States},
  author={Acemoglu, Daron and Johnson, Simon and Kermani, Amir and Kwak, James and Mitton, Todd},
  journal={Journal of Financial Economics},
  volume={121},
  number={2},
  pages={368--391},
  year={2016},
  publisher={Elsevier}
}

@article{kreif2016examination,
  title={Examination of the synthetic control method for evaluating health policies with multiple treated units},
  author={Kreif, No{\'e}mi and Grieve, Richard and Hangartner, Dominik and Turner, Alex James and Nikolova, Silviya and Sutton, Matt},
  journal={Health economics},
  volume={25},
  number={12},
  pages={1514--1528},
  year={2016},
  publisher={Wiley Online Library}
}

@article{ben2021augmented,
  title={The augmented synthetic control method},
  author={Ben-Michael, Eli and Feller, Avi and Rothstein, Jesse},
  journal={Journal of the American Statistical Association},
  number={just-accepted},
  pages={1--34},
  year={2021},
  publisher={Taylor \& Francis}
}

@article{hazlett2018trajectory,
  title={Trajectory balancing: A general reweighting approach to causal inference with time-series cross-sectional data},
  author={Hazlett, Chad and Xu, Yiqing},
  journal={Available at SSRN 3214231},
  year={2018}
}

@article{ferman2019synthetic,
  title={Synthetic controls with imperfect pre-treatment fit},
  author={Ferman, Bruno and Pinto, Cristine},
  journal={arXiv preprint arXiv:1911.08521},
  year={2019}
}

@article{chernozhukov2021exact,
  title={An exact and robust conformal inference method for counterfactual and synthetic controls},
  author={Chernozhukov, Victor and W{\"u}thrich, Kaspar and Zhu, Yinchu},
  journal={Journal of the American Statistical Association},
  pages={1--16},
  year={2021},
  publisher={Taylor \& Francis}
}

@article{bottmer2021design,
  title={A Design-Based Perspective on Synthetic Control Methods},
  author={Bottmer, Lea and Imbens, Guido and Spiess, Jann and Warnick, Merrill},
  journal={arXiv preprint arXiv:2101.09398},
  year={2021}
}

@article{cole2020impact,
  title={The impact of the Wuhan Covid-19 lockdown on air pollution and health: a machine learning and augmented synthetic control approach},
  author={Cole, Matthew A and Elliott, Robert JR and Liu, Bowen},
  journal={Environmental and Resource Economics},
  volume={76},
  number={4},
  pages={553--580},
  year={2020},
  publisher={Springer}
}

@article{hollingsworth2020tactics,
  title={Tactics for design and inference in synthetic control studies: An applied example using high-dimensional data},
  author={Hollingsworth, Alex and Wing, Coady},
  journal={Available at SSRN 3592088},
  year={2020}
}

@article{lounici2008sup,
  title={Sup-norm convergence rate and sign concentration property of Lasso and Dantzig estimators},
  author={Lounici, Karim},
  journal={Electronic Journal of statistics},
  volume={2},
  pages={90--102},
  year={2008},
  publisher={Institute of Mathematical Statistics and Bernoulli Society}
}

@article{bayat2020synthetic,
  title={Synthetic control, synthetic interventions, and COVID-19 spread: Exploring the impact of lockdown measures and herd immunity},
  author={Bayat, Niloofar and Morrin, Cody and Wang, Yuheng and Misra, Vishal},
  journal={arXiv preprint arXiv:2009.09987},
  year={2020}
}

@article{gibson2020understanding,
  title={Understanding the economic impact of COVID-19 stay-at-home orders: a synthetic control analysis},
  author={Gibson, John and Sun, Xiaojin Aaron},
  journal={Available at SSRN 3601108},
  year={2020}
}

@article{agarwal2020synthetic,
  title={Synthetic interventions},
  author={Agarwal, Anish and Shah, Devavrat and Shen, Dennis and others},
  journal={arXiv preprint arXiv:2006.07691},
  year={2020}
}

@article{shen2022heterogeneous,
  title={Heterogeneous Synthetic Learner for Panel Data},
  author={Shen, Ye and Wan, Runzhe and Cai, Hengrui and Song, Rui},
  journal={arXiv preprint arXiv:2212.14580},
  year={2022}
}

@article{abadie2021synthetic,
  title={Synthetic controls for experimental design},
  author={Abadie, Alberto and Zhao, Jinglong},
  journal={arXiv preprint arXiv:2108.02196},
  year={2021}
}
\bibliographystyle{agsm}

\newpage
\begin{center}
{\large\bf SUPPLEMENTARY MATERIAL}
\end{center}

\appendix

\setcounter{table}{0}
\setcounter{figure}{0}
\renewcommand{\thetable}{A\arabic{table}}
\renewcommand{\thefigure}{A\arabic{figure}}

In this document, we provide supplementary materials to the paper "Efficiently Learning Synthetic Control Models for High-dimensional Disaggregated Data", including additional simulation results, the real data set details and technical proofs.


\section{Additional Simulation Results} \label{sec:additionalSimu}

%
%
In this section, we first present the simulation results under Setting  (1), including MSE, ATT estimation bias and computational time  for fitting Synthetic Control weights using each method.

\begin{figure*}[!ht]
\centering
\includegraphics[width=\textwidth]{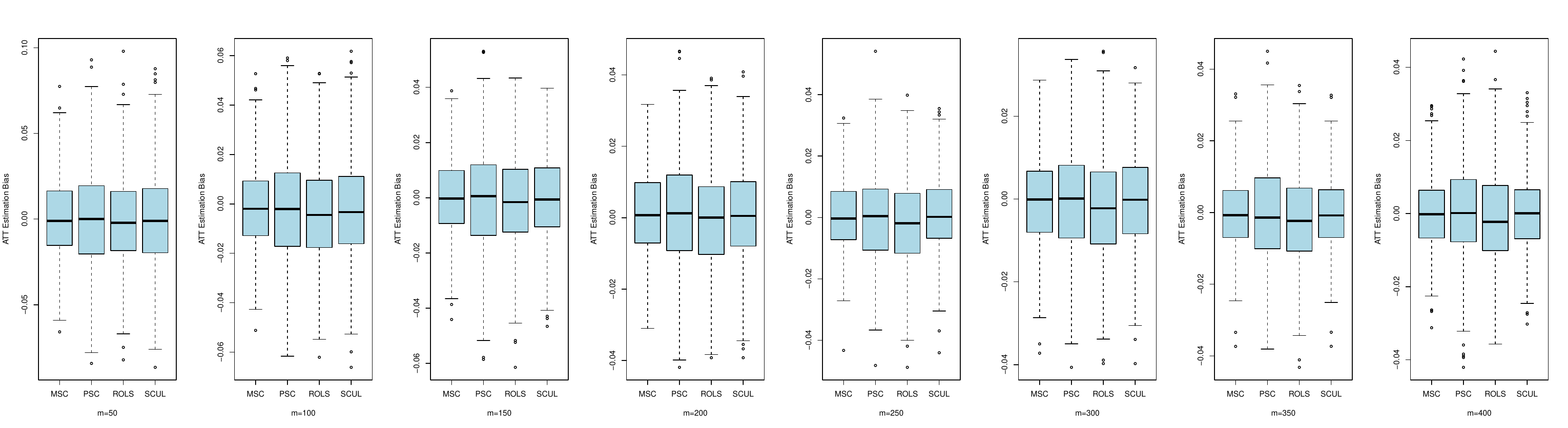}
\caption{ATT estimation bias for MSC under Setting (1) with $T_0=100$ pre-treatment periods, $T_1=10$ post-treatment periods, $m=50,150,200,250,300,350,400$ treated units, and $n=400$ control units. The experiments are repeated 500 times in total. }
\end{figure*}

\newpage

\begin{figure}[!h]
\centering
\includegraphics[width=\linewidth]{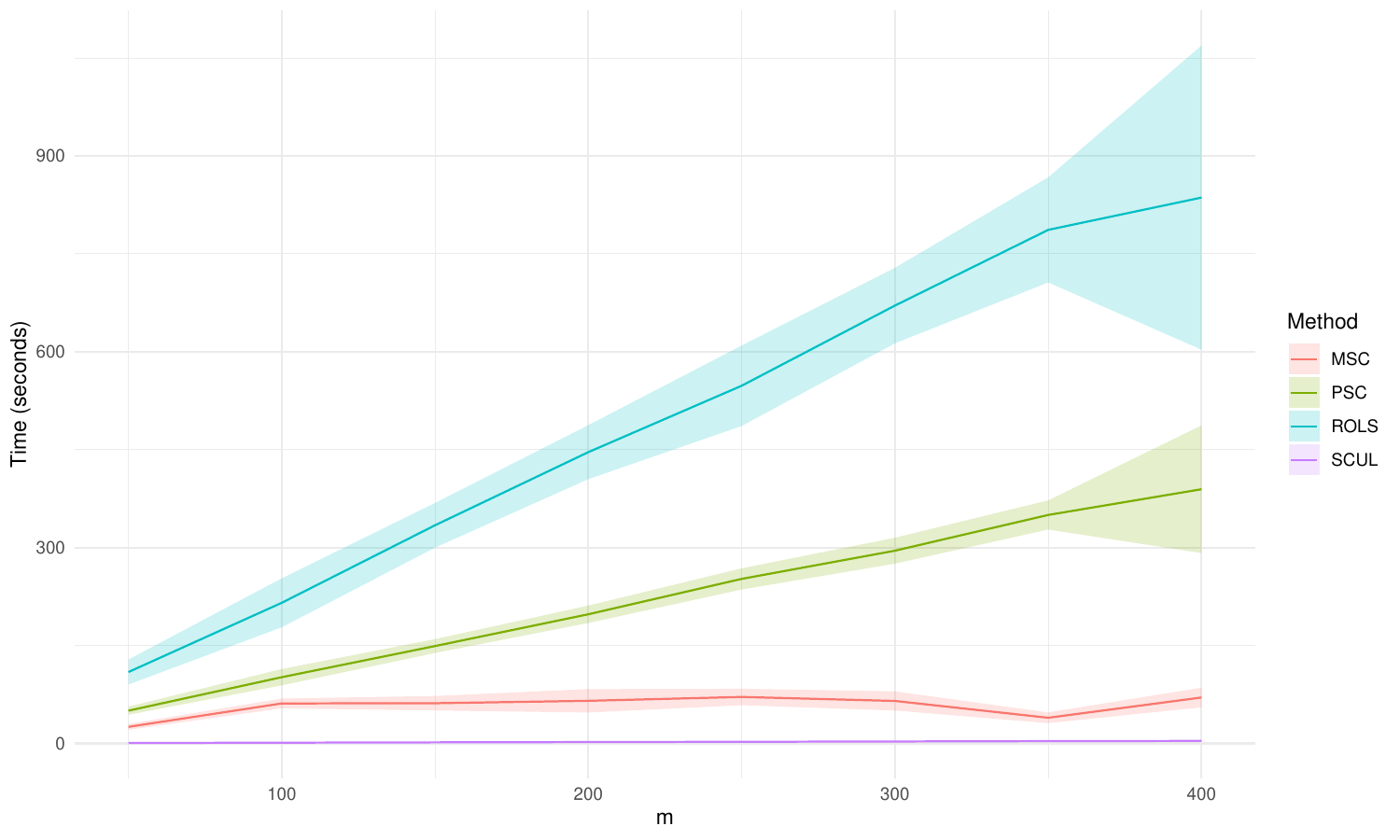}
\caption{Computational time analysis for various methods under Setting (1) with $T_0=100$ pre-treatment periods, $T_1=10$ post-treatment periods, $m=50,150,200,250,300,350,400$ treated units, and $n=400$ control units. The experiments are repeated 500 times, with the solid line representing the average and the shadow area representing one standard error.}
\end{figure}

\newpage

\begin{figure}[!h]
\centering
\includegraphics[width=0.4\linewidth]{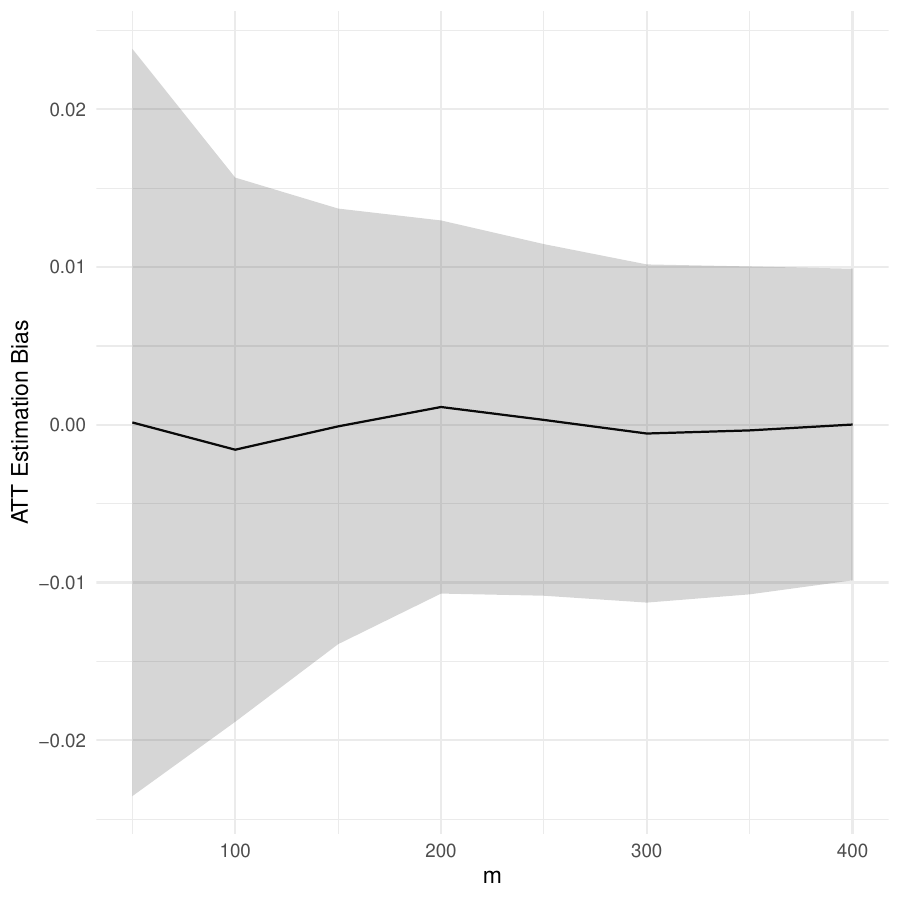}
\caption{ATT estimation bias for MSC under Setting (1) with $T_0=100$ pre-treatment periods, $T_1=10$ post-treatment periods, $m=50,150,200,250,300,350,400$ treated units, and $n=400$ control units. The experiments are repeated 500 times in total, with the solid line representing the average and the shadow area representing one standard error.}
\end{figure}

\begin{table*}[!h]
\centering
\caption{The corresponding RMSE means of independent 500 runs for fitting Synthetic Control weights under Setting (1) with $T_0=100$ pre-treatment periods, $T_1=10$ post-treatment periods, $m=50,150,200,250,300,350,400$ treated units, and $n=400$ control units.}	
\label{tb:mse1}
\begin{tabular}{|l|r|r|r|r|r|r|r|r|} \hline
 m & 50 & 100 & 150 & 200 & 250 & 300 & 350 & 400 \\ \hline
MSC & 0.48 & 0.48 & 0.48 & 0.48 & 0.48 & 0.48 & 0.48 & 0.48 \\ \hline
  PSC & 0.48 & 0.48 & 0.48 & 0.48 & 0.48 & 0.48 & 0.48 & 0.48 \\ \hline
  ROLS & 0.46 & 0.46 & 0.46 & 0.46 & 0.46 & 0.46 & 0.46 & 0.46 \\ \hline
  SCUL & 2.00 & 2.01 & 2.01 & 2.01 & 2.01 & 2.01 & 2.02 & 2.02 \\ \hline
\end{tabular}
\end{table*}

\newpage
In the simulation, the penalty parameters are predetermined through cross-validation and we set $\lambda = 0.03$ for MSC. In the following, we present a sensitivity analysis of the parameter tuning for MSC. 

\begin{figure*}[!ht]
\centering
\includegraphics[width=\textwidth]{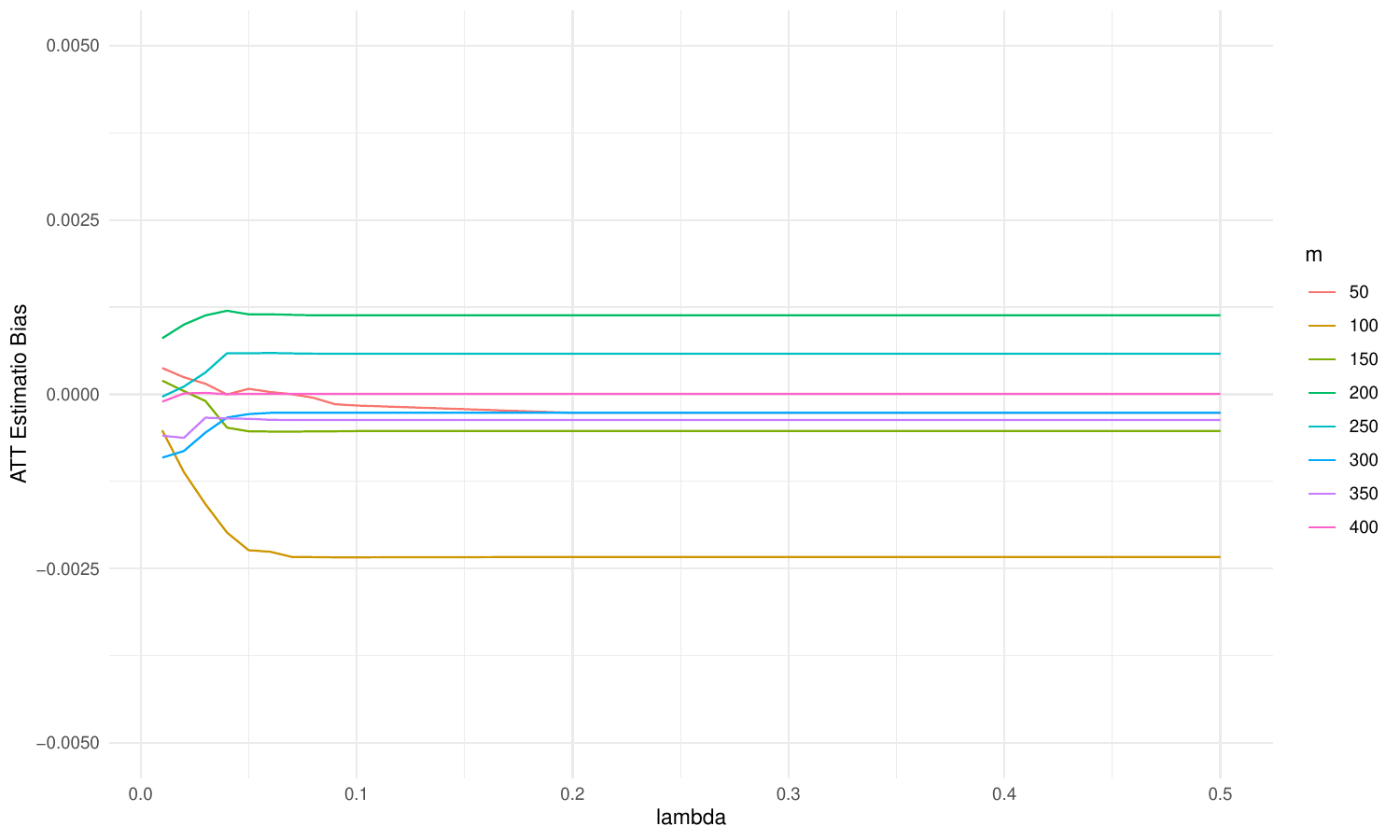}
\caption{Mean ATT estimation bias for MSC under Setting (1) with $T_0=100$ pre-treatment periods, $T_1=10$ post-treatment periods, $m=50,150,200,250,300,350,400$ treated units, $n=400$ control units and $\lambda = 0.01, 0.02, 0.03, 0.04, 0.05, 0.06, 0.07, 0.08, 0.09, 0.10, 0.10, 0.20, 0.30, 0.40, 0.50$. The experiments are repeated 500 times. }
\end{figure*}

\newpage

\begin{figure*}[!ht]
\centering
\includegraphics[width=\textwidth]{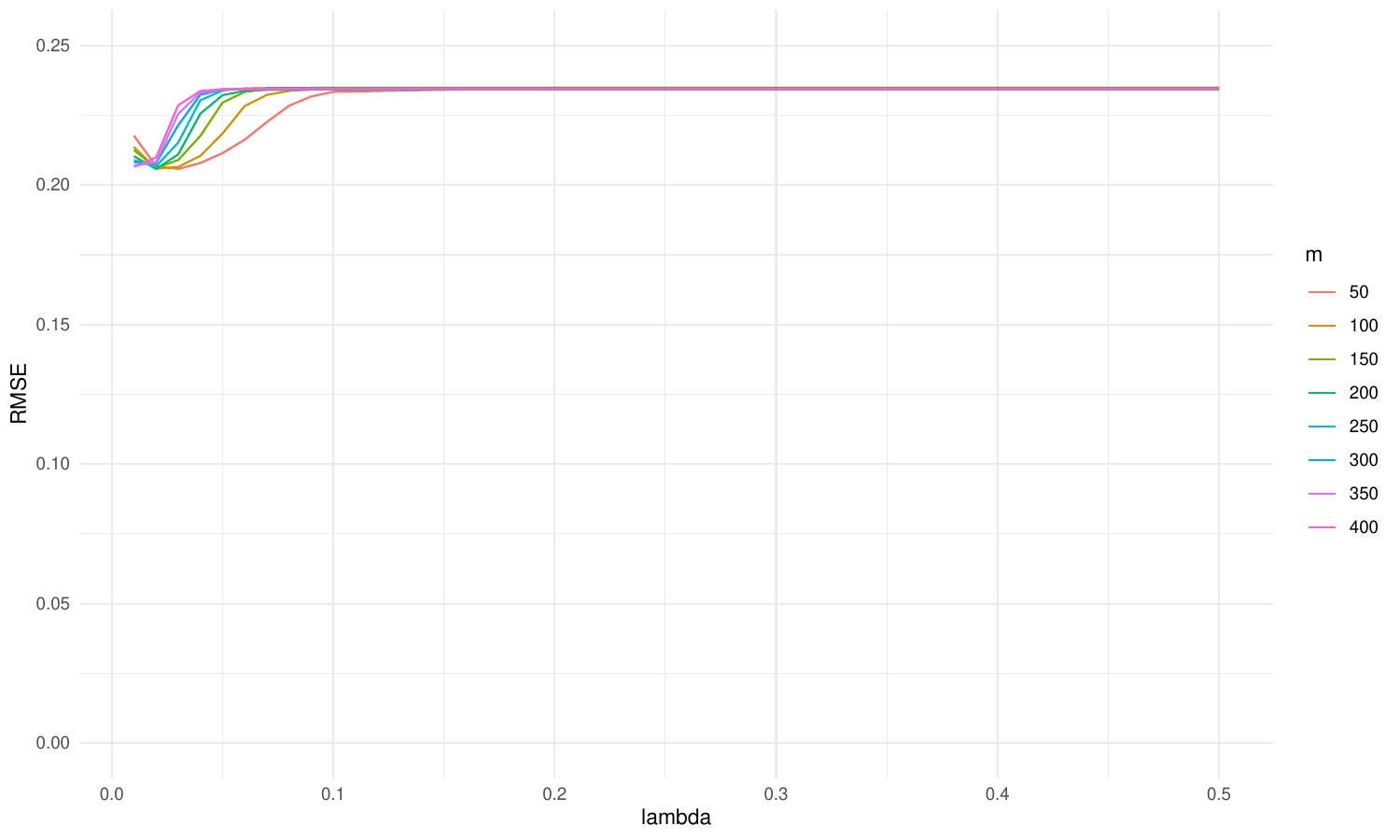}
\caption{Mean RMSE for MSC under Setting (1) with $T_0=100$ pre-treatment periods, $T_1=10$ post-treatment periods, $m=50,150,200,250,300,350,400$ treated units, $n=400$ control units and $\lambda = 0.01, 0.02, 0.03, 0.04, 0.05, 0.06, 0.07, 0.08, 0.09, 0.10, 0.10, 0.20, 0.30, 0.40, 0.50$. The experiments are repeated 500 times. }
\end{figure*}

\newpage
\begin{figure*}[!ht]
\centering
\includegraphics[width=\textwidth]{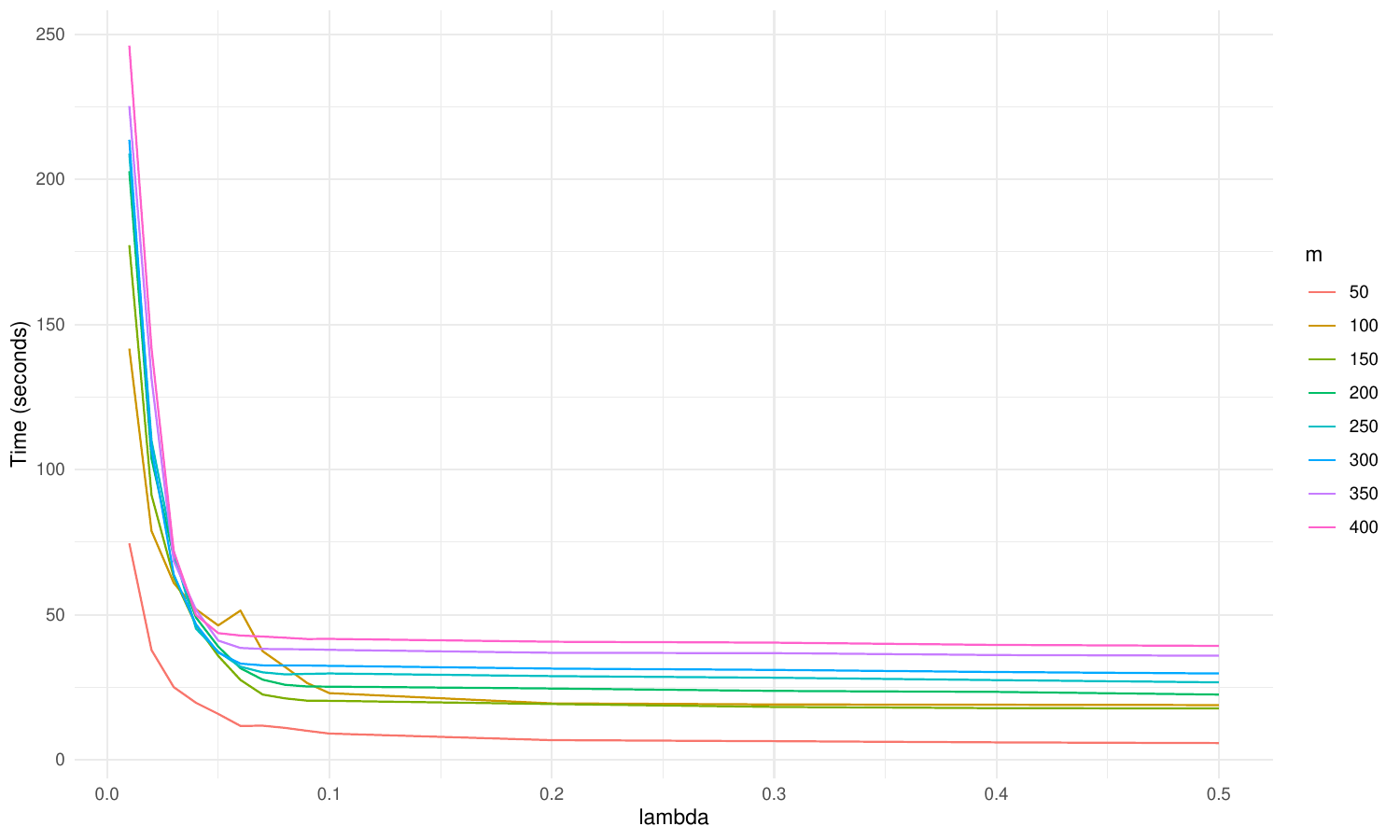}
\caption{Mean computational time for MSC under Setting (1) with $T_0=100$ pre-treatment periods, $T_1=10$ post-treatment periods, $m=50,150,200,250,300,350,400$ treated units, $n=400$ control units and $\lambda = 0.01, 0.02, 0.03, 0.04, 0.05, 0.06, 0.07, 0.08, 0.09, 0.10, 0.10, 0.20, 0.30, 0.40, 0.50$. The experiments are repeated 500 times. }
\end{figure*}

\newpage
\begin{figure*}[!ht]
\centering
\includegraphics[width=\textwidth]{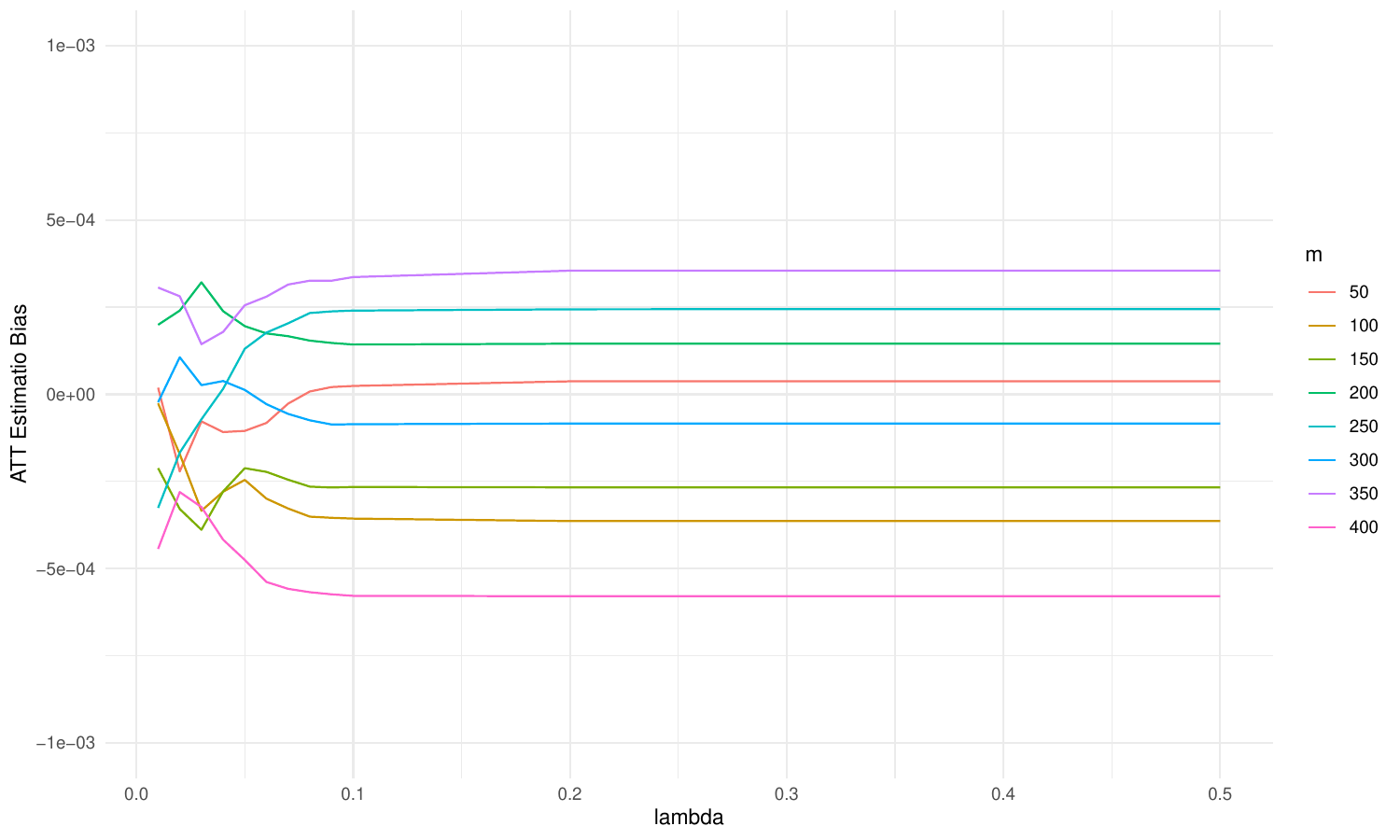}
\caption{Mean ATT estimation bias for MSC under Setting (2) with $T_0=100$ pre-treatment periods, $T_1=10$ post-treatment periods, $m=50,150,200,250,300,350,400$ treated units, $n=400$ control units, $s=1000$ and $\lambda = 0.01, 0.02, 0.03, 0.04, 0.05, 0.06, 0.07, 0.08, 0.09, 0.10, 0.10, 0.20, 0.30, 0.40, 0.50$. The experiments are repeated 500 times. }
\end{figure*}

\newpage
\begin{figure*}[!ht]
\centering
\includegraphics[width=\textwidth]{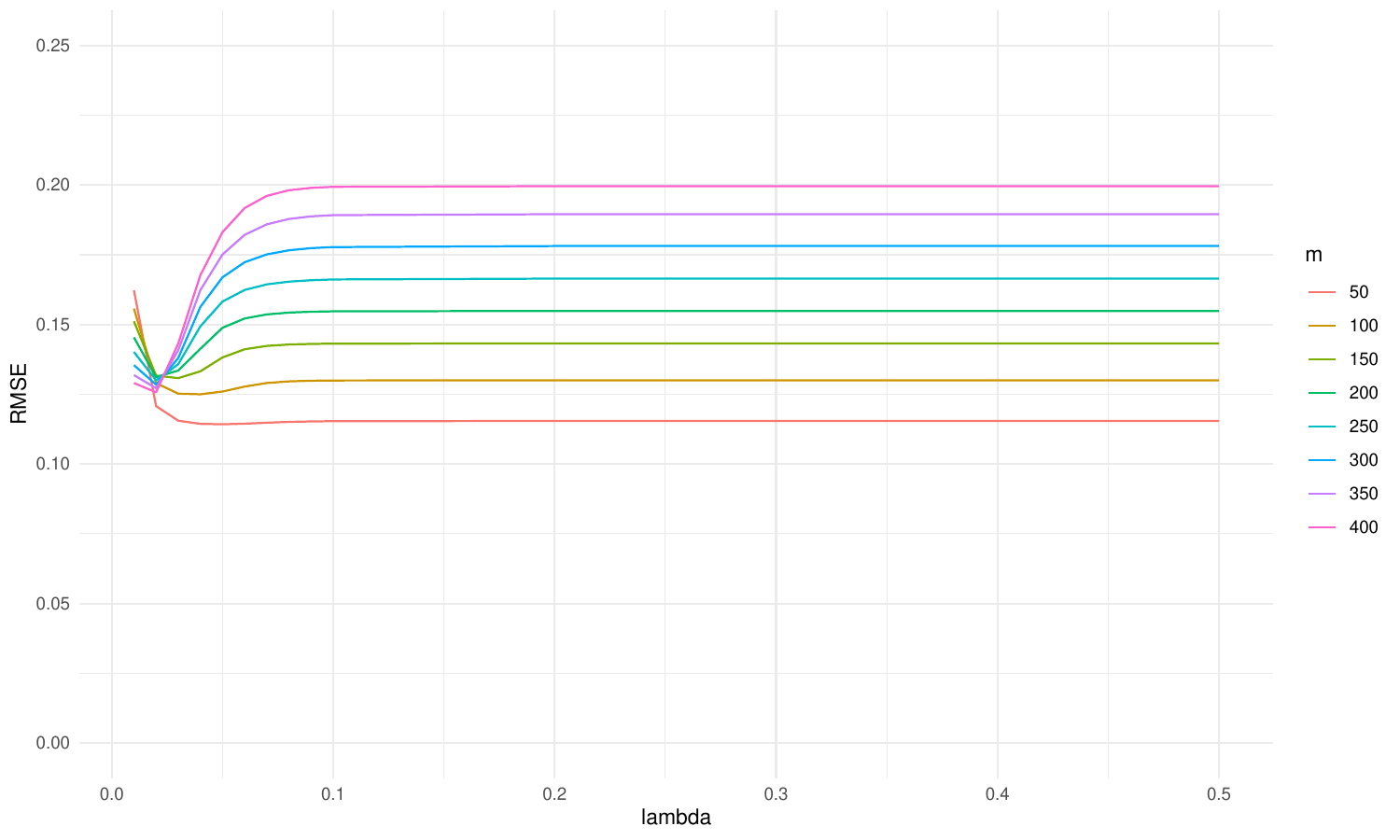}
\caption{Mean RMSE for MSC under Setting (2) with $T_0=100$ pre-treatment periods, $T_1=10$ post-treatment periods, $m=50,150,200,250,300,350,400$ treated units, $n=400$ control units, $s=1000$  and $\lambda = 0.01, 0.02, 0.03, 0.04, 0.05, 0.06, 0.07, 0.08, 0.09, 0.10, 0.10, 0.20, 0.30, 0.40, 0.50$. The experiments are repeated 500 times. }
\end{figure*}

\newpage
\begin{figure*}[!ht]
\centering
\includegraphics[width=\textwidth]{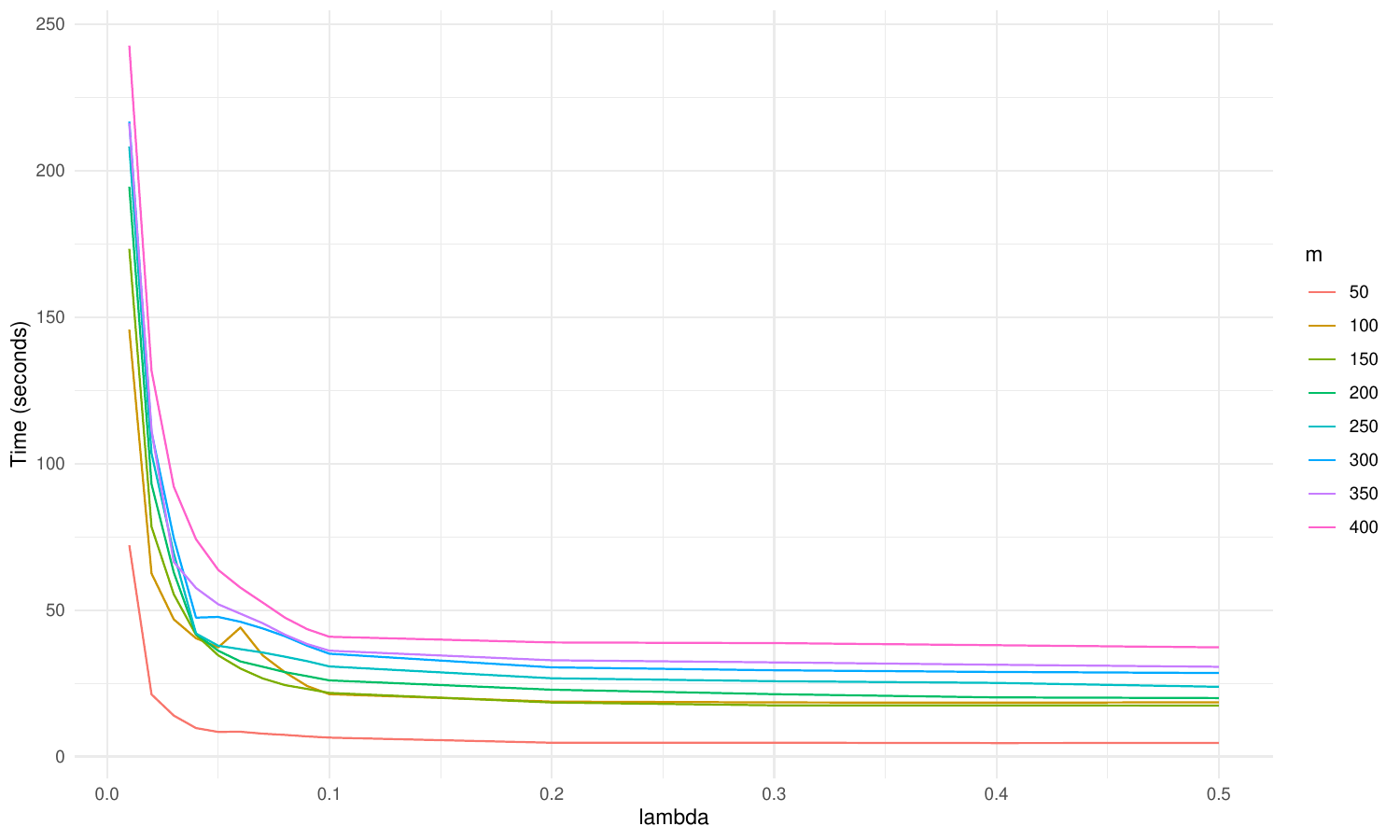}
\caption{Mean computational time for MSC under Setting (2) with $T_0=100$ pre-treatment periods, $T_1=10$ post-treatment periods, $m=50,150,200,250,300,350,400$ treated units, $n=400$ control units, $s=1000$  and $\lambda = 0.01, 0.02, 0.03, 0.04, 0.05, 0.06, 0.07, 0.08, 0.09, 0.10, 0.10, 0.20, 0.30, 0.40, 0.50$. The experiments are repeated 500 times. }
\end{figure*}

\newpage
\section{Real Data Set Details}

In this section, we provide the Stay-at-Home Orders implementation details used in our real data application.

\begin{figure}[ht]
\centering
\includegraphics[width=\linewidth]{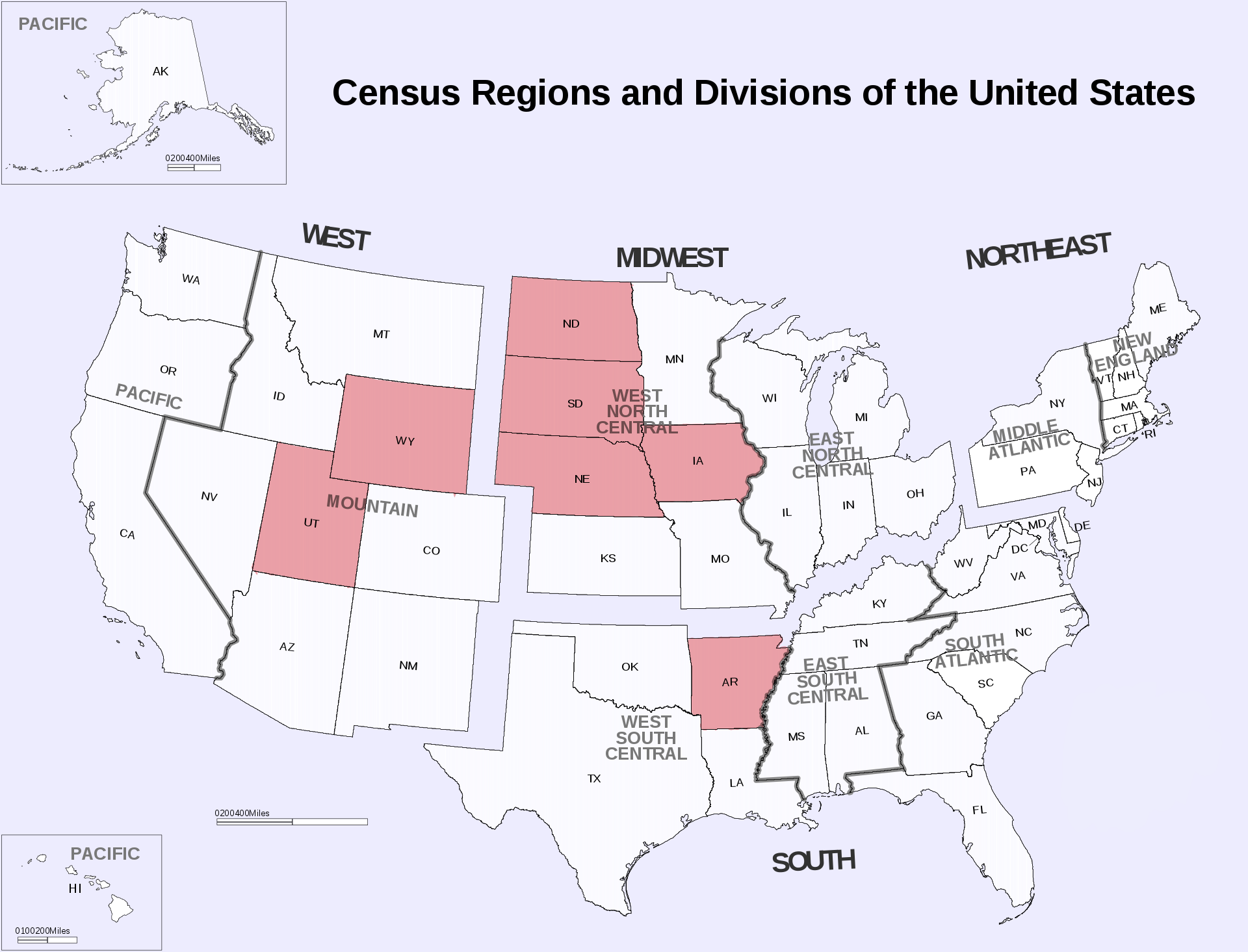}
\caption{The location of states without COVID-19 Stay-at-Home Orders. }
\label{fig:map}
\end{figure}

\newpage

\begin{table}[!h]
\centering
\caption{Statewide stay-at-home orders in response to COVID-19 (Table 1 in \cite{gibson2020understanding})}
\label{tb:SAHtime}
 \scalebox{0.8}{\begin{tabular}{llll}
\hline \hline State & Order start date & State & Order start date \\
\hline Alabama & April 4, 2020 & Montana & March 28, 2020 \\
Alaska & March 28, 2020 & Nebraska & None \\
Arizona & March 31, 2020 & Nevada & April 1, 2020 \\
Arkansas & None & New Hampshire & March 27, 2020 \\
California & March 19, 2020 & New Jersey & March 21, 2020 \\
Colorado & March 26, 2020 & New Mexico & March 24, 2020 \\
Connecticut & March 23, 2020 & New York & March 20, 2020 \\
Delaware & March 24, 2020 & North Carolina & March 30, 2020 \\
Florida & April 2, 2020 & North Dakota & None \\
Georgia & April 3, 2020 & Ohio & March 23, 2020 \\
Hawaii & March 25, 2020 & Oklahoma & April 1, 2020 \\
Idaho & March 25, 2020 & Oregon & March 23, 2020 \\
Illinois & March 21, 2020 & Pennsylvania & April 1, 2020 \\
Indiana & March 24, 2020 & Rhode Island & March 28, 2020 \\
Iowa & None & South Carolina & April 7, 2020 \\
Kansas & March 30, 2020 & South Dakota & None \\
Kentucky & March 26, 2020 & Tennessee & March 31, 2020 \\
Louisiana & March 23, 2020 & Texas & April 2, 2020 \\
Maine & April 2, 2020 & Utah & None \\
Maryland & March 30, 2020 & Vermont & March 24, 2020 \\
Massachusetts & March 24, 2020 & Virginia & March 30, 2020 \\
Michigan & March 24, 2020 & Washington & March 24, 2020 \\
Minnesota & March 27, 2020 & West Virginia & March 24, 2020 \\
Mississippi & April 3, 2020 & Wisconsin & March 25, 2020 \\
Missouri & April 6, 2020 & Wyoming & None \\
\hline \hline
\end{tabular}}
 \end{table}

\newpage

\begin{figure}[ht]
\centering
\includegraphics[width= 0.55 \linewidth]{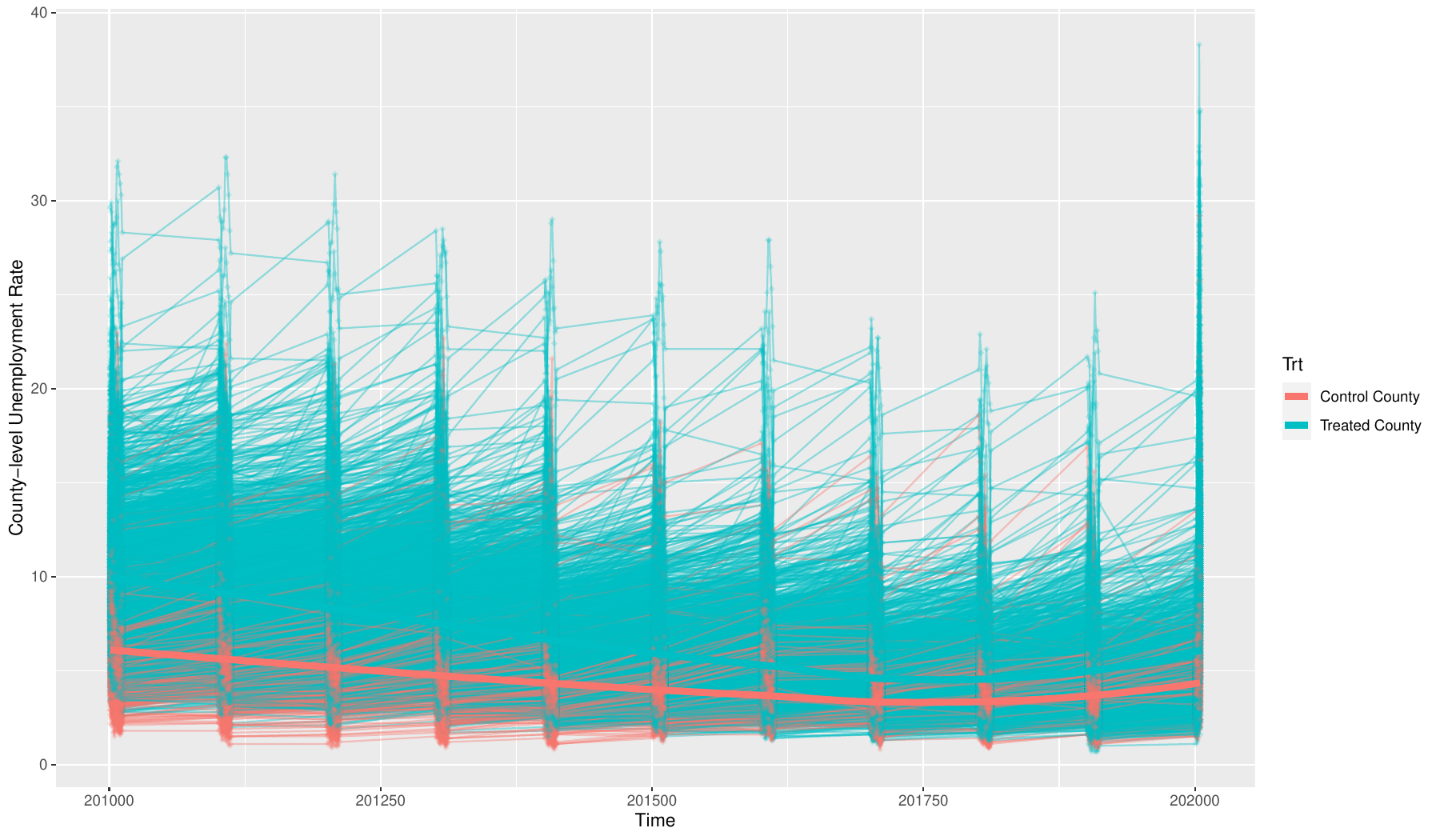}
\caption{Spaghetti plot of the county-level unemployment rate.}
\label{fig:spaghetti}
\end{figure}
\begin{figure}[h]
\centering
\includegraphics[width= 0.55 \linewidth]{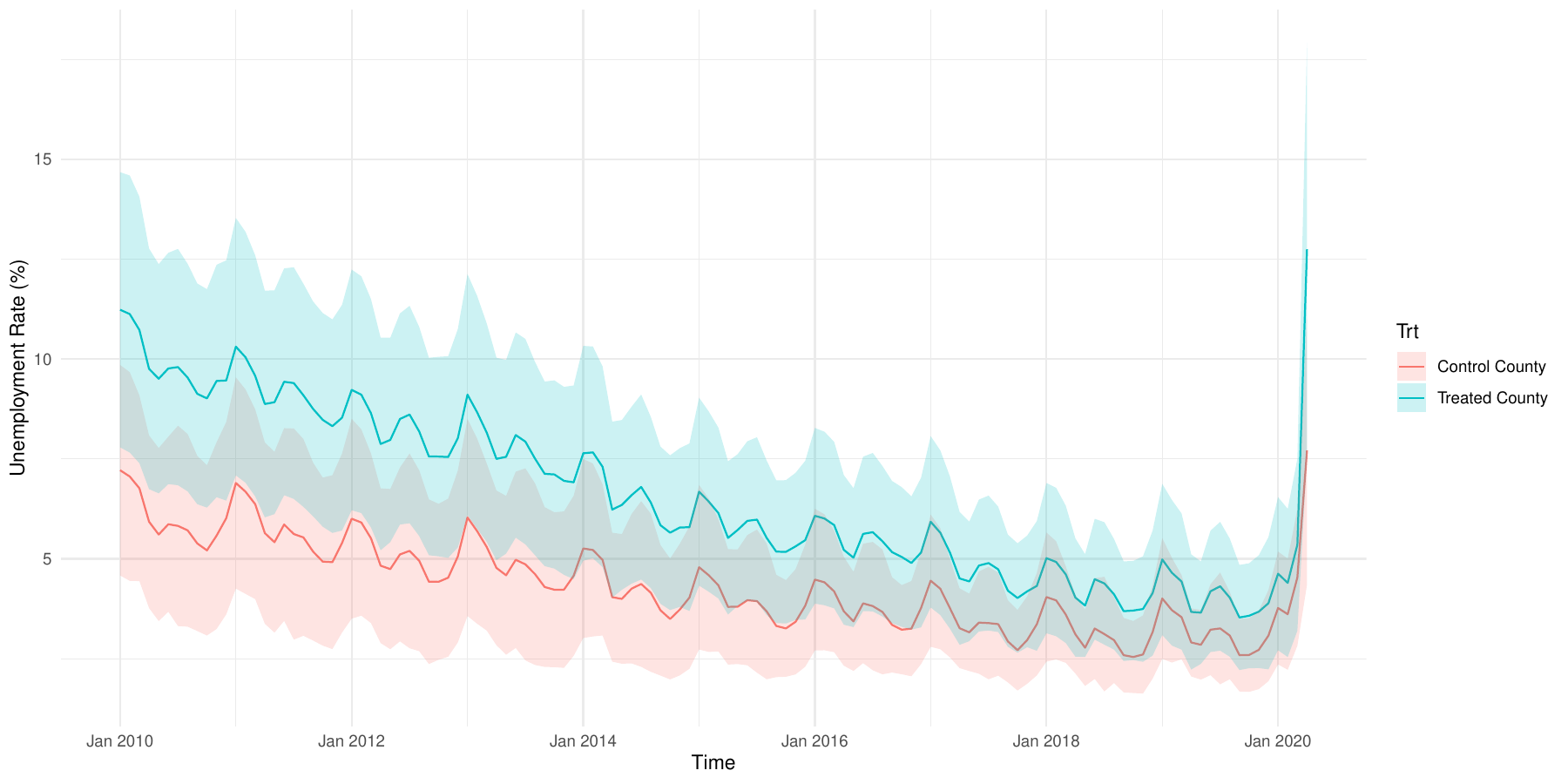}
\caption{The county-level unemployment rate, with the solid line representing the average and the shadow area representing one standard error. }
\label{fig:realdatamean}
\end{figure}

\newpage

\section{Proofs}\label{sec:proof}
In this section, we provide detailed proofs of the results in the main paper. We start by introducing additional notations used in the proofs:
\begin{itemize}
	\item For a vector $\mathbf{Z}$, denote $\|\mathbf{Z}\|_2 $ as the Euclidean norm. 
\item Denote  $\partial f(\mathbf{\Theta}^{*})$ as  the subgradient of $f(\mathbf{\Theta}) $ at $\mathbf{\Theta}^{*}$.
\item For a regularizer  $g(\cdot)$, denote the dual form of $g(\cdot)$ as 
\begin{equation*}
\tilde{g}(\mathbf{V}) = sup_{\mathbf{U} \neq \mathbf{0}} \left< \mathbf{U},\mathbf{V} \right> /g(\mathbf{U})
\end{equation*}
for matrix $\mathbf{U}$ and $\mathbf{V}$ with commensurate dimensions. As studied in \cite{negahban2012unified}, the dual form of  $g_1(\mathbf{\Theta}) = \sum_{i=1}^{n} \sum_{j=1}^{m} |\mathbf{\Theta}_{i,j}| $ is $\tilde{g}_1(\mathbf{\Theta}) = \max_{i,j}|\mathbf{\Theta}_{i,j}| = \|\mathbf{\Theta}\|_{\max}$. 
\item For any subspace $\mathcal{M}$ and regularizer $g(\cdot)$, define  the subspace compatibility constant with respect to the pair $(g(\cdot),\|\cdot\|_{F})$ by 
\begin{equation*}
\Psi(\mathcal{M}):=\sup _{u \in \mathcal{M} \backslash\{\mathbf{0}\}} g(u)/\|u\|_{F}.
\end{equation*}
\end{itemize}


\subsection{Proof of Theorem \ref{thm:ErrorBound}}

We will begin by proving a more general version of Theorem \ref{thm:ErrorBound}. In this general setting, we consider an optimization problem of the form:
\begin{equation} \label{eq:general_Object}
\underset{\mathbf{\Theta} \in \mathbb{R}^{n \times m}}{\arg \min } \mathcal{\mathcal{L}(\mathbf{\Theta})} = f(\mathbf{\Theta}) +\lambda g \left( \mathbf{\Theta}\right) \text{with } f(\mathbf{\Theta}) = \frac{1}{\sqrt{T_0}}\|\mathbf{Y}-\mathbf{X}  \mathbf{\Theta}\|_*,
\end{equation}
where $f(\cdot)$ is a convex function, and $g(\cdot)$ is a regularizer satisfying the following conditions: 
\begin{enumerate}
	\item (Triangle Inequality)  $g(\theta+\gamma) \leq g(\theta)+ g(\gamma)$ for $\forall$ $\theta, \gamma$ in the domain of $g$;
	\item (Absolute Homogeneity) $g(s\theta) = |s|g(\theta)$ for all scalars $s$ and $\gamma$ in the domain of $g$.;
	\item (Decomposable) Given subspaces $\mathcal{M}$ and its orthogonal complement \begin{equation*}\overline{\mathcal{M}}^{\perp}:=\left\{\gamma \mid \langle \gamma, \theta \rangle=0 \text { for all }\theta \in \overline{\mathcal{M}}\right\}, g(\theta+\gamma) = g(\theta)+ g(\gamma) \end{equation*} for all $\theta \in \mathcal{M}$ and $\gamma \in \overline{\mathcal{M}}^{\perp}$.
\end{enumerate}
It's important to note that these conditions are quite general and are satisfied by commonly used regularizers like the $\mathcal{L}_1$ norm, weighted $\mathcal{L}_1$ norm, Group Lasso, and nuclear norm, as demonstrated by \cite{negahban2012unified}.

Now, let's state the theorem:

\begin{thm}{(Estimation Error)}  \label{thm:GeneralErrorBound}
For the error matrix $\mathbf{E}_{T_0\times m}$ , denote
\begin{equation*}
\Lambda: =\left\{ \mathbf{Z}:\mathbf{Z}\in \mathcal{R}^{T_0\times m}, \|\mathbf{Z}\|_2 \leq 1, \mathbf{U_{\mathbf{E}}}^{\top} \mathbf{Z}=0, \mathbf{Z} \mathbf{V}_{\mathbf{E}}=0\right\}.
\end{equation*}
 Then for any fixed constant $c>1$, $\lambda \geq \frac{c}{\sqrt{T_0}}  \left\{  \tilde{g}\left(\mathbf{X}^{\top} \mathbf{U}_{\mathbf{E}} \mathbf{V}_{\mathbf{E}}^{\top}\right) + sup_{ \mathbf{Z} \in \Lambda} \tilde{g}\left(\mathbf{X}^{\top} \mathbf{Z}\right)\right\}$, and any regularizer $g(\cdot)$ satisfying the above Triangle Inequality, Absolute Homogeneity and Decomposable conditions, with Assumption  \ref{ass:noAnticipation}, \ref{ass:Consistency},  \ref{ass:sparse}, and \ref{ass:RSC} hold, we have
\begin{equation} 
	\|\widehat{\mathbf{\Theta}}-\mathbf{\Theta}^*\|_F \leq  \frac{(c+1)\lambda \Psi(\overline{\mathcal{S}})}{c \phi_{\mathbf{E}, g}(\mathcal{S}, c)} .
\end{equation}
\end{thm}
This theorem establishes the estimation error in our general optimization problem. In the following, we aim to provide a rigorous proof.

\textbf{Proof:} Firstly, note that by Lemma \ref{lemma:subgradient}, we have
\begin{equation*}
sup \tilde{g}\left(\partial f(\mathbf{\Theta}^{*})\right) \leq \frac{1}{\sqrt{T_0}}  \left\{  \tilde{g}\left(\mathbf{X}^{\top} \mathbf{U}_{\mathbf{E}} \mathbf{V}_{\mathbf{E}}^{\top}\right) + sup_{ \mathbf{Z} \in \Lambda} \tilde{g}\left(\mathbf{X}^{\top} \mathbf{Z}\right)\right\},
\end{equation*}
which implies that $\lambda \geq  c \operatorname{sup} \tilde{g}\left(\partial f(\mathbf{\Theta}^{*})\right)$.

Then by Lemma \ref{lemma:cg}, for any decomposable regularizer $g$, since $\lambda \geq c \operatorname{sup} \tilde{g}\left(\partial f(\mathbf{\Theta}^{*})\right)$, we have that  $\mathbf{\Delta}=\widehat{\mathbf{\Theta}}-\mathbf{\Theta}^*$ belongs to the set
\begin{equation*}
\mathcal{C}_g(\mathcal{S}, c)=\left\{\mathbf{\Delta} \in \mathbb{R}^{n  \times m}: g\left(\mathbf{\Delta}_{\overline{\mathcal{S}}^{\perp}}\right) \leq \frac{c+1}{c-1} g\left(\mathbf{\Delta}_{\overline{\mathcal{S}}}\right)\right\}.
\end{equation*}

Thus by lemma \ref{lemma:FNormInequl} we have
\begin{equation} \label{eq:FNorm}
\phi\|\mathbf{\Delta}\|_F^2 \leq \frac{1}{\sqrt{T_0}}\left\|\mathbf{Y}-\mathbf{X}\left(\mathbf{\Theta}^*+\mathbf{\Delta}\right)\right\|_*-\frac{1}{\sqrt{T_0}}\left\|\mathbf{Y}-\mathbf{X} \mathbf{\Theta}^*\right\|_* +\frac{1}{\sqrt{T_0}} \left| \operatorname{tr}\left(\mathbf{\Delta}^{\top} \mathbf{X}^{\top} \mathbf{U}_{\mathbf{E}} \mathbf{V}_{\mathbf{E}}^{\top}\right) \right|.
\end{equation}

Since $\widehat{\mathbf{\Theta}}$ is optimal for optimization problem \eqref{eq:general_Object} and $\mathbf{\Theta}^{*}$ is feasible,
\begin{equation*}
\frac{1}{\sqrt{T_0}}\|\mathbf{Y}-\mathbf{X} \widehat{\mathbf{\Theta}}\|_*+ \lambda g\left( \widehat{\mathbf{\Theta}}\right) \leq \frac{1}{\sqrt{T_0}}\|\mathbf{Y}-\mathbf{X} \mathbf{\Theta}^{*}\|_*+ \lambda g\left(\mathbf{\Theta}^{*}\right) ,
\end{equation*}
i.e.
\begin{equation} \label{eq:feasible}
\frac{1}{\sqrt{T_0}}\|\mathbf{Y}-\mathbf{X} \left(\mathbf{\Theta}^*+\mathbf{\Delta}\right)\|_* - \frac{1}{\sqrt{T_0}}\|\mathbf{Y}-\mathbf{X} \mathbf{\Theta}^{*}\|_* \leq  \lambda  \left\{ g\left(\mathbf{\Theta}^{*}\right)  -  g\left( \widehat{\mathbf{\Theta}}\right)  \right\} .
\end{equation}
Combining the above equation with Equation \eqref{eq:FNorm} fields
\begin{equation}\label{eq:FNormInequl}
\phi_{\mathbf{E}, g}(\mathcal{S}, c)\|\mathbf{\Delta}\|_F^2 \leq \frac{1}{\sqrt{T_0}} \operatorname{tr}\left(\mathbf{\Delta}^{\top} \mathbf{X}^{\top} \mathbf{U}_{\mathbf{E}} \mathbf{V}_{\mathbf{E}}^{\top}\right) + \lambda  \left\{ g\left(\mathbf{\Theta}^{*}\right)  -  g\left( \widehat{\mathbf{\Theta}}\right)  \right\} .
\end{equation}
Observe that:
\begin{equation*}
 \frac{1}{\sqrt{T_0}} \operatorname{tr}\left(\mathbf{\Delta}^{\top} \mathbf{X}^{\top} \mathbf{U}_{\mathbf{E}} \mathbf{V}_{\mathbf{E}}^{\top}\right) =  \frac{1}{\sqrt{T_0}} \left< \mathbf{X}^{\top} \mathbf{U}_{\mathbf{E}} \mathbf{V}_{\mathbf{E}}^{\top}, \mathbf{\Delta} \right>  \leq  \frac{1}{\sqrt{T_0}} \tilde{g}\left( \mathbf{X}^{\top} \mathbf{U}_{\mathbf{E}} \mathbf{V}_{\mathbf{E}}^{\top} \right) g\left(  \mathbf{\Delta} \right),
\end{equation*}
where $\frac{1}{\sqrt{T_0}} \tilde{g}\left( \mathbf{X}^{\top} \mathbf{U}_{\mathbf{E}} \mathbf{V}_{\mathbf{E}}^{\top} \right) \leq \frac{1}{\sqrt{T_0}} \tilde{g}\left( \mathbf{X}^{\top} \mathbf{U}_{\mathbf{E}} \mathbf{V}_{\mathbf{E}}^{\top} \right) +\frac{1}{\sqrt{T_0}} sup_{ \mathbf{Z} \in \Lambda} \tilde{g}\left(\mathbf{X}^{\top} \mathbf{Z}\right) \leq \frac{\lambda}{c}$. Thus it follows that
\begin{equation} \label{eq:trInequal}
	 \frac{1}{\sqrt{T_0}} \operatorname{tr}\left(\mathbf{\Delta}^{\top} \mathbf{X}^{\top} \mathbf{U}_{\mathbf{E}} \mathbf{V}_{\mathbf{E}}^{\top}\right)  \leq  \frac{1}{\sqrt{T_0}} \tilde{g}\left( \mathbf{X}^{\top} \mathbf{U}_{\mathbf{E}} \mathbf{V}_{\mathbf{E}}^{\top} \right) g\left(  \mathbf{\Delta} \right) \leq \frac{\lambda}{c}  g\left(  \mathbf{\Delta} \right)   =\frac{\lambda}{c} \left\{ g\left(\mathbf{\Delta}_{\overline{\mathcal{S}}}\right) +g\left(\mathbf{\Delta}_{\overline{\mathcal{S}}^{\perp}}\right)\right\} .
\end{equation}
As per Lemma \ref{lemma:decomposableG}:
\begin{equation} \label{eq:penInequal}
 g\left(\mathbf{\Theta}^{*}\right)  -  g\left( \widehat{\mathbf{\Theta}}\right) \leq g\left(\mathbf{\Delta}_{\overline{\mathcal{S}}}\right) - g\left(\mathbf{\Delta}_{\overline{\mathcal{S}}^{\perp}}\right).
\end{equation}

Therefore, combining Equation \eqref{eq:FNormInequl} with  Equation \eqref{eq:trInequal} and Equation \eqref{eq:penInequal} ,  we find that
\begin{equation*}
\begin{aligned}
\phi_{\mathbf{E}, g}(\mathcal{S}, c)\|\mathbf{\Delta}\|_F^2  & \leq \frac{\lambda}{c}\left\{ g\left(\mathbf{\Delta}_{\overline{\mathcal{S}}}\right) +g\left(\mathbf{\Delta}_{\overline{\mathcal{S}}^{\perp}}\right)\right\}  +\lambda \left\{ g\left(\mathbf{\Delta}_{\overline{\mathcal{S}}}\right) -g\left(\mathbf{\Delta}_{\overline{\mathcal{S}}^{\perp}}\right)\right\} \\
  & = \lambda \left\{\frac{c+1}{c}  g\left(\mathbf{\Delta}_{\overline{\mathcal{S}}}\right)  - \frac{c-1}{c}g\left(\mathbf{\Delta}_{\overline{\mathcal{S}}^{\perp}}\right)\right\} \\
  & \leq   \lambda  \frac{c+1}{c}  g\left(\mathbf{\Delta}_{\overline{\mathcal{S}}}\right) \\
  &\leq  \lambda  \frac{c+1}{c}  \Psi(\overline{\mathcal{S}}) \|\mathbf{\Delta}\|_F,
\end{aligned}
\end{equation*}
which follows
\begin{equation*}
	\|\mathbf{\Delta}\|_F \leq  \frac{(c+1)\lambda \Psi(\overline{\mathcal{S}})}{c \phi_{\mathbf{E}, g}(\mathcal{S}, c)} .
\end{equation*}
Then the proof of Theorem \ref{thm:GeneralErrorBound} is completed.

\textbf{Now we are able to prove Theorem \ref{thm:ErrorBound}. }In this paper, we are interested in the Lasso penalty $g(\mathbf{\Theta}) = \sum_{i=1}^{n} \sum_{j=1}^{m} |\mathbf{\Theta}_{i,j}| = \|\mathbf{\Theta} \|_{1}$, with the dual form  $\tilde{g}(\mathbf{\Theta}) = \|\mathbf{\Theta} \|_{\max}$.  Note that by \cite{negahban2012unified}, we have $\Psi(\overline{\mathcal{S}}) \leq \sqrt{s}$. Thus Theorem \ref{thm:ErrorBound} holds.

\subsection{Proof of  Corollary \ref{coro}}
 To prove Theorem Corollary \ref{coro}, it suffices to prove that 
 \begin{equation*}
 	2 \left\{\frac{n \log(nT_0)}{T_0} \right\}^{1/4} \geq \frac{1}{\sqrt{T_0}}  \left\{  \tilde{g}\left(\mathbf{X}^{\top} \mathbf{U}_{\mathbf{E}} \mathbf{V}_{\mathbf{E}}^{\top}\right) + sup_{ \mathbf{Z} \in \Lambda} \tilde{g}\left(\mathbf{X}^{\top} \mathbf{Z}\right)\right\} 
 \end{equation*}
 holds with probability greater than $1-\sqrt{2} \sigma\left\{ \frac{\log(nT_0)}{nT_0 } \right\}^{1/4}$.

Note that for $\forall$ $\mathbf{Z} \in \Lambda$, we have
\begin{equation*}
\tilde{g}\left(\mathbf{X}^{\top} \mathbf{Z}\right) = \|\mathbf{X}^{\top} \mathbf{Z}\|_{\max} \leq  \|\mathbf{X} ^{\top}\|_{\max} \| \mathbf{Z}\|_{\max} \leq  \|\mathbf{X} ^{\top}\|_{\max} \| \mathbf{Z}\|_{2} \leq  \|\mathbf{X} \|_{\max}.
\end{equation*} Also notice that 
\begin{equation*}
\tilde{g}\left(\mathbf{X}^{\top} \mathbf{U}_{\mathbf{E}} \mathbf{V}_{\mathbf{E}}^{\top}\right)  = \|\mathbf{X}^{\top}\mathbf{U}_{\mathbf{E}} \mathbf{V}_{\mathbf{E}}^{\top}\|_{\max} \leq \|\mathbf{X}^{\top}\|_{\max}  \|  \mathbf{U}_{\mathbf{E}} \|_{\max}  \|  \mathbf{V}_{\mathbf{E}}\|_{\max} \leq \|\mathbf{X}\|_{\max},
\end{equation*}
where the last inequality follows from the fact that $\|  \mathbf{V}\|_{\max} \leq \|  \mathbf{V}\|_{2}  = 1$ for any orthogonal matrice $\mathbf{V}$. Therefore, we have
\begin{equation*}
\frac{1}{\sqrt{T_0}}  \left\{  \tilde{g}\left(\mathbf{X}^{\top} \mathbf{U}_{\mathbf{E}} \mathbf{V}_{\mathbf{E}}^{\top}\right) + sup_{ \mathbf{Z} \in \Lambda} \tilde{g}\left(\mathbf{X}^{\top} \mathbf{Z}\right)\right\} \leq \frac{2}{\sqrt{T_0}}  \|\mathbf{X}\|_{\max}.
\end{equation*}
Under Assumption \ref{ass:subGaussian}, by Lemma \ref{lemma:subGaussian}, we have
\begin{equation*}
\begin{aligned}
P\left( \|\mathbf{X}\|_{\max} > \left\{nT_0 \log(nT_0)\right\}^{1/4}\right) &  \leq \frac{\sigma \sqrt{2\log(nT_0)} }{ \left\{nT_0 \log(nT_0)\right\}^{1/4} + \Mean \left\{ \|\mathbf{X}_{i,t}\| \right\}}   \\
& \leq \frac{\sigma \sqrt{2\log(nT_0)} }{ \left\{nT_0 \log(nT_0)\right\}^{1/4}} \\
& =  \sqrt{2} \sigma\left\{ \frac{\log(nT_0)}{nT_0 } \right\}^{1/4}.
\end{aligned} 
\end{equation*}
Therefore, with probability greater than $1-\sqrt{2} \sigma\left\{ \frac{\log(nT_0)}{nT_0 } \right\}^{1/4}$, we have
\begin{equation*}
\frac{1}{\sqrt{T_0}}  \left\{  \tilde{g}\left(\mathbf{X}^{\top} \mathbf{U}_{\mathbf{E}} \mathbf{V}_{\mathbf{E}}^{\top}\right) + sup_{ \mathbf{Z} \in \Lambda} \tilde{g}\left(\mathbf{X}^{\top} \mathbf{Z}\right)\right\}\leq \frac{2}{\sqrt{T_0}}  \|\mathbf{X}\|_{\max}  \leq  \frac{2}{\sqrt{T_0}} \left\{nT_0 \log(nT_0)\right\}^{1/4} = 2 \left\{\frac{n \log(nT_0)}{T_0} \right\}^{1/4}.
\end{equation*}
Then the proof is completed.

\subsection{Proof of Theorem \ref{thm:ATT}}

In Theorem \ref{thm:ATT}, our objective is to establish an error bound for the ATT estimator $\widehat{\delta}$, defined as:
\begin{equation*}
\widehat{\delta}   =  \frac{1}{m}\sum_{i=1}^{m} \left\{ Y_{i,T_0+1} - \widehat{Y}_{i,T_0+1}(0) \right\} .
\end{equation*}
Note that 
\begin{equation*}
\widehat{\delta} - \delta  =  \frac{1}{m}\sum_{i=1}^{m} \left\{ Y_{i,T_0+1} - \widehat{Y}_{i,T_0+1}(0)\right\} -  \frac{1}{m}\sum_{i=1}^{m} \left\{ Y_{i,T_0+1} - Y_{i,T_0+1}(0)\right\} =\frac{1}{m} \sum_{j=1}^{m} \left\{ Y_{i,T_0+1}(0)- \widehat{Y}_{i,T_0+1}(0)\right\}.
\end{equation*}
Recall that $\mathbf{Y}_{post} = \left(Y_{m+1, T_0+1},Y_{m+2, T_0+1}, \cdots, Y_{m+n, T_0+1}\right)$. Now, let's denote $\Gamma \triangleq \mathbf{Y}_{post} \left( \widehat{\mathbf{\Theta}}-\mathbf{\Theta}^* \right) $. This quantity follows:
\begin{equation} \label{eq:ProofATT1}
\widehat{\delta} - \delta = \frac {\sum_{i=1}^{m} \Gamma_{i} }{m} \leq \sqrt{\frac{\sum_{i=1}^{m}\Gamma_{i}^2}{m} } = \frac{\|\Gamma\|_F}{\sqrt{m}}  =  \frac{\|\mathbf{Y}_{post} \left( \widehat{\mathbf{\Theta}}-\mathbf{\Theta}^* \right) \|_F}{\sqrt{m}} \leq  \frac{\|\mathbf{Y}_{post}  \|_F \|\widehat{\mathbf{\Theta}}-\mathbf{\Theta}^* \|_F}{\sqrt{m}},
\end{equation}
where the first inequality corresponds to the AM-QM inequality, and the second inequality corresponds to the Cauchy-Schwarz inequality.

Recall that by Assumption \ref{ass:subGaussian}, the potential outcome $Y_{i, t}(0)$ is $\sigma$-sub Gaussian random variable, which implies 
\begin{equation*}
\Var \left( Y_{i, T_0+1}^2 \right) \leq \sigma^2.
\end{equation*}
Recall that $\Mean \left\{  Y_{i, T_0+1}\right\} \leq L$, thus
\begin{equation*}
\Mean\left( Y_{i, T_0+1}^2 \right)  = \Var \left( Y_{i, T_0+1}^2 \right) + \left[ \Mean \left\{  Y_{i, T_0+1}\right\} \right]^2 \leq \sigma^2+L^2.
\end{equation*}
Note that
\begin{equation*}
\|\mathbf{Y}_{post}  \|_F= \sqrt{\sum_{i=m}^{m+n} Y_{i, T_0+1}^2}.
\end{equation*}
Thus by Markov inequality, for any $a >0$ ,
\begin{equation*}
P\left( \|\mathbf{Y}_{post}  \|_F > a \right) = P\left( \sum_{i=m}^{m+n} Y_{i, T_0+1}^2 > a^2 \right) \leq  \frac{\sum_{i=m}^{m+n} \Mean\left( Y_{i, T_0+1}^2 \right)}{a^2} \leq \frac{n \left( \sigma^2+L^2\right)}{a^2}.
\end{equation*} 
Thus Equation \eqref{eq:ProofATT1} can be further expressed as:
\begin{equation*}
\widehat{\delta} - \delta  \leq  \frac{(c+1) \sqrt{s} \lambda  a }{c \phi_{\mathbf{E}, g}(\mathcal{S}, c) \sqrt{m}},
\end{equation*}
with probability greater than $  1-\sqrt{2} \sigma\left\{ \log(nT_0)/(nT_0) \right\}^{1/4} -  \frac{n\left( \sigma^2+L^2\right)}{a^2}$.

Choose 
$a = \frac{n^{1/2}T_0^{1/8}}{\{log(T_0)\}^{1/8}}$
Then
\begin{equation*}
\widehat{\delta} - \delta  \leq   \left\{ \frac{T_0}{log(T_0)}\right\} ^{1/8}\frac{(c+1) \sqrt{ns} \lambda }{c \phi_{\mathbf{E}, g}(\mathcal{S}, c)\sqrt{m}} 
 \end{equation*}
with probability greater than $  1-\sqrt{2} \sigma\left\{ \log(nT_0)/(nT_0) \right\}^{1/4} -  \left( \sigma^2+L^2\right) \left\{ \frac{log(T_0)}{T_0}\right\} ^{1/8} $.

\subsection{Auxiliary Results}

\begin{lemma} \label{lemma:subgradient} For  any  regularizer $g$ satisfying the Triangle Inequality, we have 
\begin{equation*}
sup \tilde{g}\left(\partial f(\mathbf{\Theta}^{*})\right) \leq \frac{1}{\sqrt{T_0}}  \left\{  \tilde{g}\left(\mathbf{X}^{\top} \mathbf{U}_{\mathbf{E}} \mathbf{V}_{\mathbf{E}}^{\top}\right) + sup_{ \mathbf{Z} \in \Lambda} \tilde{g}\left(\mathbf{X}^{\top} \mathbf{Z}\right)\right\}.
\end{equation*}
\end{lemma}
\textbf{Proof:} Lemma 11 in \cite{molstad2021new} establishes the subgradient $\partial f(\mathbf{\Theta}^{*})$, which characterizes the change in the objective function $f$ around the optimal solution $\mathbf{\Theta}^{*}$, expressed as follows
\begin{equation*}
\partial f(\mathbf{\Theta}^{*}) =\left\{-\frac{1}{\sqrt{T_0}}\mathbf{X}^{\top} \mathbf{U}_{\mathbf{E}} \mathbf{V}_{\mathbf{E}}^{\top}-\frac{1}{\sqrt{T_0}}\mathbf{X}^{\top} \mathbf{Z}:\mathbf{Z}\in \mathcal{R}^{T_0\times m}, \|\mathbf{Z}\|_2 \leq 1, \mathbf{U_{\mathbf{E}}}^{\top} \mathbf{Z}=0, \mathbf{Z} \mathbf{V}_{\mathbf{E}}=0\right\}.
\end{equation*}
Denote
\begin{equation*}
\Lambda: =\left\{ \mathbf{Z}:\mathbf{Z}\in \mathcal{R}^{T_0\times m}, \|\mathbf{Z}\|_2 \leq 1, \mathbf{U_{\mathbf{E}}}^{\top} \mathbf{Z}=0, \mathbf{Z} \mathbf{V}_{\mathbf{E}}=0\right\},
\end{equation*}
we can then express the subgradient as:
\begin{equation*}
\begin{aligned}
 sup \tilde{g}\left(\partial f(\mathbf{\Theta}^{*})\right) & =  sup_{ \mathbf{Z} \in \Lambda} \tilde{g}\left(\frac{1}{\sqrt{T_0}}\mathbf{X}^{\top} \mathbf{U}_{\mathbf{E}} \mathbf{V}_{\mathbf{E}}^{\top}+\frac{1}{\sqrt{T_0}}\mathbf{X}^{\top} \mathbf{Z}\right) = \frac{1}{\sqrt{T_0}}sup_{ \mathbf{Z} \in \Lambda} \tilde{g}\left(\mathbf{X}^{\top} \mathbf{U}_{\mathbf{E}} \mathbf{V}_{\mathbf{E}}^{\top}+ \mathbf{X}^{\top} \mathbf{Z}\right)
\end{aligned}
\end{equation*}
By the Triangle Inequality, we have
\begin{equation*}
\tilde{g}\left(\mathbf{X}^{\top} \mathbf{U}_{\mathbf{E}} \mathbf{V}_{\mathbf{E}}^{\top}+ \mathbf{X}^{\top} \mathbf{Z}\right) \leq \tilde{g}\left(\mathbf{X}^{\top} \mathbf{U}_{\mathbf{E}} \mathbf{V}_{\mathbf{E}}^{\top} \right) + \tilde{g}\left(\mathbf{X}^{\top} \mathbf{Z}\right).
\end{equation*}
Thus, the subgradient satisfies
\begin{equation*}
sup \tilde{g}\left(\partial f(\mathbf{\Theta}^{*})\right) \leq \frac{1}{\sqrt{T_0}}  \left\{  \tilde{g}\left(\mathbf{X}^{\top} \mathbf{U}_{\mathbf{E}} \mathbf{V}_{\mathbf{E}}^{\top}\right) + sup_{ \mathbf{Z} \in \Lambda} \tilde{g}\left(\mathbf{X}^{\top} \mathbf{Z}\right)\right\}.
\end{equation*}

\begin{lemma} \label{lemma:cg}
For any decomposable regularizer $g$, if $\lambda \geq c \operatorname{sup} \tilde{g}\left(\partial f(\mathbf{\Theta}^{*})\right)$, with $\partial f(\mathbf{\Theta}^{*})$ being the subgradient of a convex function $f(\mathbf{\Theta}) $ at $\mathbf{\Theta}^* \in \mathcal{S}$, , then $\mathbf{\Delta}=\widehat{\mathbf{\Theta}}-\mathbf{\Theta}^*$ belongs to the set
\begin{equation*}
\mathcal{C}_g(\mathcal{S}, c)=\left\{\mathbf{\Delta} \in \mathbb{R}^{n  \times m}: g\left(\mathbf{\Delta}_{\overline{\mathcal{S}}^{\perp}}\right) \leq \frac{c+1}{c-1} g\left(\mathbf{\Delta}_{\overline{\mathcal{S}}}\right)\right\}.
\end{equation*}
\end{lemma}
\textbf{Proof:} This lemma is a generalization of Lemma 1 in \cite{negahban2012unified} to high-dimentsional case. We notice that  $f(\cdot)$ is a convex function, thus for  $\forall$ $\mathbf{\Delta}\in \mathcal{R}^{n\times m}$,  we have 
\begin{equation} \label{eq:lemmaB2.1}
f(\mathbf{\Theta}^{*} + \mathbf{\Delta}) - f(\mathbf{\Theta}^{*}) \geq  
\left <\partial f(\mathbf{\Theta}^{*}), \mathbf{\Delta}\right> \geq -\left|  \left <\partial f(\mathbf{\Theta}^{*}), \mathbf{\Delta}\right>\right|,
\end{equation}
where $\partial f(\mathbf{\Theta}^{*})$ is the subgradient of $f(\mathbf{\Theta}) $ at $\mathbf{\Theta}^{*}$.

Recall the definition of dual form, we have 
\begin{equation*}
\left|  \left <\partial f(\mathbf{\Theta})^{*}, \mathbf{\Delta}\right>\right| \leq \tilde{g}\left(\partial f(\mathbf{\Theta}^{*})\right) g(\mathbf{\Delta}).
\end{equation*}
Note that  $\lambda \geq c \operatorname{sup} \tilde{g}\left(\partial f(\mathbf{\Theta}^{*})\right)$ and $g(\mathbf{\Delta}) = g\left(\mathbf{\Delta}_{\overline{\mathcal{S}}^{\perp}}\right) + g\left(\mathbf{\Delta}_{\overline{\mathcal{S}}}\right)$ since  $g$ is decomposable, thus the above equation can be further expressed as
\begin{equation*}
\left|  \left <\partial f(\mathbf{\Theta})^{*}, \mathbf{\Delta}\right>\right| \leq \tilde{g}\left(\partial f(\mathbf{\Theta}^{*})\right) g(\mathbf{\Delta}) \leq \frac{\lambda}{c} g(\mathbf{\Delta}) = \frac{\lambda}{c}\left\{ g\left(\mathbf{\Delta}_{\overline{\mathcal{S}}^{\perp}}\right) + g\left(\mathbf{\Delta}_{\overline{\mathcal{S}}}\right) \right\}.
\end{equation*}
Hence, Equation \eqref{eq:lemmaB2.1} can also be stated as
\begin{equation*}
 f( \widehat{\mathbf{\Theta}}) - f(\mathbf{\Theta}^{*})= f(\mathbf{\Theta}^{*} + \mathbf{\Delta}) - f(\mathbf{\Theta}^{*}) \geq   - \frac{\lambda}{c} \left\{ g\left(\mathbf{\Delta}_{\overline{\mathcal{S}}^{\perp}}\right) + g\left(\mathbf{\Delta}_{\overline{\mathcal{S}}}\right) \right\}.
\end{equation*}
Note that, by lemma \ref{lemma:decomposableG}, we have $  g(\widehat{\mathbf{\Theta}}) - g(\mathbf{\Theta}^{*})  =g(\mathbf{\Theta}^{*}+\mathbf{\Delta}) - g(\mathbf{\Theta}^{*}) \geq g\left(\mathbf{\Delta}_{\overline{\mathcal{S}}^{\perp}}\right) - g\left(\mathbf{\Delta}_{\overline{\mathcal{S}}}\right)$, which follows
\begin{equation*}
\begin{aligned}
\mathcal{L}( \widehat{\mathbf{\Theta}}) - \mathcal{L}(\mathbf{\Theta}^{*})&  = f( \widehat{\mathbf{\Theta}})  - f(\mathbf{\Theta}^{*})+ \lambda   \left\{ g( \widehat{\mathbf{\Theta}}) - g(\mathbf{\Theta}^{*}) \right\} \\
& \geq  - \frac{\lambda}{c}\left\{ g\left(\mathbf{\Delta}_{\overline{\mathcal{S}}^{\perp}}\right) + g\left(\mathbf{\Delta}_{\overline{\mathcal{S}}}\right) \right\} +  \lambda \left\{ g\left(\mathbf{\Delta}_{\overline{\mathcal{S}}^{\perp}}\right) - g\left(\mathbf{\Delta}_{\overline{\mathcal{S}}}\right) \right\} \\
& = \lambda \left\{\left(1-\frac{1}{c}\right) g\left(\mathbf{\Delta}_{\overline{\mathcal{S}}^{\perp}}\right) -\left(1+\frac{1}{c}\right) g\left(\mathbf{\Delta}_{\overline{\mathcal{S}}}\right) \right\}.
\end{aligned}
\end{equation*}
On the other hand, since $\widehat{\mathbf{\Theta}}$ is optimal for optimization problem \eqref{eq:general_Object} and $\mathbf{\Theta}^{*}$ is feasible, we have $\mathcal{L}( \widehat{\mathbf{\Theta}}) - \mathcal{L}(\mathbf{\Theta}^{*}) \leq 0$. Therefore we have
\begin{equation*}
0 \geq \lambda \left\{\left(1-\frac{1}{c}\right) g\left(\mathbf{\Delta}_{\overline{\mathcal{S}}^{\perp}}\right) -\left(1+\frac{1}{c}\right) g\left(\mathbf{\Delta}_{\overline{\mathcal{S}}}\right) \right\},
\end{equation*}
which follows $g\left(\mathbf{\Delta}_{\overline{\mathcal{S}}^{\perp}}\right) \leq \frac{1+1/c}{1-1/c} g\left(\mathbf{\Delta}_{\overline{\mathcal{S}}}\right) = \frac{c+1}{c-1} g\left(\mathbf{\Delta}_{\overline{\mathcal{S}}}\right)$. Hence the claim holds.

\begin{lemma}{(Lemma 13 in \cite{molstad2021new})} \label{lemma:FNormInequl}
For all $\mathbf{\Delta} \in \mathcal{C}_g(\mathcal{S}, c)$,
\begin{equation*}
 \phi\|\mathbf{\Delta}\|_F^2 \leq \frac{1}{\sqrt{T_0}}\left\|\mathbf{Y}-\mathbf{X}\left(\mathbf{\Theta}^*+\mathbf{\Delta}\right)\right\|_*-\frac{1}{\sqrt{T_0}}\left\|\mathbf{Y}-\mathbf{X} \mathbf{\Theta}^*\right\|_* +\frac{1}{\sqrt{T_0}} \left| \operatorname{tr}\left(\mathbf{\Delta}^{\top} \mathbf{X}^{\top} \mathbf{U}_{\mathbf{E}} \mathbf{V}_{\mathbf{E}}^{\top}\right) \right|
\end{equation*}
\end{lemma}

\begin{lemma}\label{lemma:decomposableG}
For $\forall$ $\mathbf{\Theta}\in \mathcal{R}^{n\times m}$,  any decomposable regularizer $g(\cdot)$ and $\mathbf{\Delta}$ $\in \mathcal{R}^{n\times m}$, we have 
\begin{equation*}
g\left( \mathbf{\Theta} + \mathbf{\Delta}\right) -  g\left(\mathbf{\Theta}\right)  \geq g\left(\mathbf{\Delta}_{\overline{\mathcal{S}}^{\perp}}\right) - g\left(\mathbf{\Delta}_{\overline{\mathcal{S}}}\right).
\end{equation*}
\end{lemma}

\textbf{Proof:}
Since $g$ is a decomposable regularizer, we have 
\begin{equation} \label{eq:lemma4.1}
g\left( \mathbf{\Theta} + \mathbf{\Delta}\right)   = g\left( \mathbf{\Theta} + \mathbf{\Delta}_{\overline{\mathcal{S}}}+ \mathbf{\Delta}_{\overline{\mathcal{S}}^{\perp}}\right)  = g\left( \mathbf{\Theta} + \mathbf{\Delta}_{\overline{\mathcal{S}}}\right)  + g\left(\mathbf{\Delta}_{\overline{\mathcal{S}}^{\perp}}\right).
\end{equation}
Note that by the Triangle Inequality, 
\begin{equation*}
	g\left( \mathbf{\Theta} + \mathbf{\Delta}_{\overline{\mathcal{S}}}\right)+ g\left( - \mathbf{\Delta}_{\overline{\mathcal{S}}}\right) \geq  g\left( \mathbf{\Theta} \right),
\end{equation*}
which implies 
\begin{equation*}
g\left( \mathbf{\Theta} + \mathbf{\Delta}_{\overline{\mathcal{S}}}\right) \geq  g\left( \mathbf{\Theta} \right)  - g\left(-\mathbf{\Delta}_{\overline{\mathcal{S}}}\right) = g\left( \mathbf{\Theta} \right)  - g\left(\mathbf{\Delta}_{\overline{\mathcal{S}}}\right).
\end{equation*}
Therefore,

Hence Equation \eqref{eq:lemma4.1} can be further expressed as
\begin{equation*}
g\left( \mathbf{\Theta} + \mathbf{\Delta}\right) -  g\left(\mathbf{\Theta}\right)   \geq  g\left( \mathbf{\Theta} \right)  - g\left(\mathbf{\Delta}_{\overline{\mathcal{S}}}\right)+g\left(\mathbf{\Delta}_{\overline{\mathcal{S}}^{\perp}}\right) - g\left( \mathbf{\Theta} \right)  \geq g\left(\mathbf{\Delta}_{\overline{\mathcal{S}}^{\perp}}\right) - g\left(\mathbf{\Delta}_{\overline{\mathcal{S}}}\right).
\end{equation*}

\begin{lemma} \label{lemma:subGaussian}
 Let $\{X_i\}_{i=1}^{n}$ be  $\sigma$-sub gaussian random variables with mean $\mu$ (not necessarily independent). Then for any $b>0$, we have
\begin{equation*}
P\left( \max |X_i|- \mu> b\right) \leq \frac{\sigma \sqrt{2\log(n)} }{b}.
\end{equation*}
\end{lemma}
\textbf{Proof:} Without loss of generality, assume $\mu=0$. For any $a>0$, we have
\begin{equation*}
e^{a\mathbb{E}\left\{\max |X_i|\right\}} \leq \mathbb{E}\left[e^{a\max |X_i|}\right]= \mathbb{E}\left[\max e^{a |X_i|}\right] \leq \mathbb{E}\left[\sum_{i=1}^n e^{a |X_i|}\right] =  \sum_{i=1}^n\mathbb{E}\left[ e^{a |X_i|}\right]\leq n e^{\frac{a^2 \sigma^2}{2}},
\end{equation*}
where the first inequality is by Jensen's inequality and the last inequality is by the definition of $\sigma$-sub gaussian.  Therefore, 
\begin{equation*}
\mathbb{E}\left\{\max |X_i|\right\} \leq \frac{\log(n)}{a} + \frac{a \sigma^2}{2}
\end{equation*}
for any $a>0$. Hence,
\begin{equation*}
\mathbb{E}\left\{\max |X_i|\right\} \leq \inf_{a>0}\left\{\frac{\log(n)}{a} + \frac{a \sigma^2}{2} \right\} = \sigma \sqrt{2\log(n)}.
\end{equation*}
Then for any $b>0$, by Markov's Inequality, we have
\begin{equation*}
P\left( \max |X_i| > b\right) \leq \frac{\mathbb{E}\left\{\max |X_i|\right\} }{b} \leq \frac{\sigma \sqrt{2\log(n)} }{b}.
\end{equation*}

\end{document}